\documentclass[traditabstract]{aa} 
\usepackage{graphicx}
\usepackage{subfig}
\usepackage{natbib}
\usepackage{txfonts}

\begin{document}

\title{On detectability of Zeeman broadening in optical spectra of
  F- and G-dwarfs \thanks{Based on observations collected at the
    European Southern Observatory, La Silla} }

  \author{Richard I. Anderson\inst{1,2}
    \and
    Ansgar Reiners\inst{2}
    \and
    Sami K. Solanki\inst{3,4}
  }

  \institute{
    Observatoire de Gen\`eve, Universit\'e de Gen\`eve,
    51 Ch. des Maillettes, CH-1290 Sauverny, Switzerland
    \and
    Institut f\"ur Astrophysik, Georg-August-Universit\"at
    G\"ottingen, Friedrich-Hund-Platz 1, D-37077 G\"ottingen,
    Germany
    \and    Max-Planck-Institut f\"ur Sonnensystemforschung,
    Max-Planck-Stra\ss e 2, D-37191 Katlenburg-Lindau, Germany
    \and
    School of Space Research, Kyung Hee University, Yongin, Gyeonggi 446-701,
    Korea\\
    \email{richard.anderson@unige.ch}
  }

  \date{Received 12 April 2010 / Accepted 11 August 2010}

  \abstract{We investigate the detectability of Zeeman broadening in
    optical Stokes~I spectra of slowly rotating sun-like stars.  To
    this end, we apply the LTE spectral line inversion package
    \texttt{SPINOR} to very-high quality CES data and explore how fit
    quality depends on the average magnetic field, $Bf$.
    One-component (OC) and two-component (TC) models are adopted.
    In OC models, the entire surface is assumed to be magnetic.
    Under this assumption, we determine formal $3\sigma$ upper limits 
    on the average magnetic field of $200$\,G for
    the Sun, and $150$\,G for $61$\,Vir (G6V). Evidence for 
    an average magnetic field of $\sim
    500$\,G is found for $59$\,Vir (G0V), and of
    $\sim 1000$\,G for
    HD\,68456 (F6V).
    A distinction between magnetic and non-magnetic regions is made in
    TC models, while assuming a homogeneous distribution of both
    components. In our TC inversions of $59$\,Vir, we investigate
    three cases: both components have equal temperatures; warm
    magnetic regions; cool magnetic regions.
    Our TC model with equal temperatures does not
    yield significant improvement over OC inversions for $59$\,Vir.
    The resulting $Bf$ values are consistent for both.
    Fit quality is significantly improved, however, by using two
    components of different temperatures.
    The inversions for $59$\,Vir that assume different temperatures for
    the two components yield results consistent with $0 - 450$\,G
    at the formal $3\sigma$ confidence level.
    We thus find a model dependence of our analysis and demonstrate
    that the influence of an additional temperature component can
    dominate over the Zeeman broadening signature, at least in
    optical data.
    Previous comparable analyses that neglected effects due to
    multiple temperature components may be prone to the same ambiguities.
}


  \maketitle
  %

\section{Introduction}

Cool stars display a range of phenomena that are typical of magnetic
activity, such as dark spots present at the stellar surface
\citep[see ][ for a review]{2005LRSP....2....8B},
enhanced chromospheric emission in, e.g., the Ca II H and K lines
\citep[e.g., ][]{1994ssac.book.....W, 2003A&A...397..147P, 2004ApJ...614..942H},
flares \citep[e.g., ][]{2000A&A...353..987F, 2000A&A...356..627M} and coronae
\citep{2001IAUS..203..475S}.  Although these manifestations of
stellar activity and hence of the underlying magnetic field are well detected
for large samples of stars, measuring the magnetic field itself turns out to be
 difficult on most cool stars.
The reason is the complex small-scale geometry of the field,
that is amply evident on the Sun. The mixture of nearly equal
amounts of magnetic flux of opposite polarities on the stellar disk leads to
cancellation of the circular polarization signal due to the longitudinal
Zeeman effect, the most straightforward diagnostic of a stellar magnetic
field.

Rapid stellar rotation partly removes the degeneracy between opposite 
magnetic polarities that lie sufficiently far apart on the stellar 
surface and thereby enables detection of a net (circular) 
polarization signal. Consequently, the magnetic field can be mapped 
on the stellar surface from the variation of
the spectral shape and amplitude of this polarization signal over a stellar
rotation, with the help of the Zeeman Doppler Imaging (ZDI) technique
\citep{1989A&A...225..456S}, which has been successfully employed to
reconstruct the large-scale distribution of the magnetic flux on the surfaces
of a number of stars  \citep{2008ASPC..384..156D, 2009ARA&A..47..333D}.

In spite of its success, ZDI does have some limitations.
First, unless the field is homogeneously and unipolarly distributed on
large scales, ZDI detects only some fraction of the true magnetic flux on the
stellar surface.
Second, the same is true for the measured field strength and hence
the magnetic energy density. In general, the field strength is underestimated
and consequently the magnetic energy density is even more strongly 
underestimated, since it is proportional to the square
of the field strength \citep[see][]{2009A&A...496..787R}.
Third, ZDI is limited to comparatively rapidly rotating stars. The slower a
star rotates (or rather the smaller its $v\sin i$) the lower the spatial
resolution of the stellar surface (and of the magnetic field) that can be
achieved.

Therefore, the method proposed and applied to cool stars by
\citet{1980ApJ...239..961R, 1980ApJ...236L.155R}, of
looking for changes in the line profile shape of the intensity (Stokes $I$)
due to Zeeman splitting, remains an interesting technique to complement ZDI.
In particular, it allows an estimate of the total magnetic flux as well as of
the field strength and filling factor to be obtained (and hence also of the
magnetic energy density).

Zeeman broadening in stellar spectra (ZB) was first measured by 
\citet{1971ApJ...164..309P} to determine the average surface magnetic
fields of Ap stars. \citet{1980ApJ...236L.155R} then measured ZB in cool stars
using his Fourier-based method which was taken up and extended by
other authors \citep{1984ApJ...277..640G,1982PASP...94..989M}. It was
soon realized, however, that the assumptions made in these Fourier
techniques were too crude. Thus, a shift towards
ever-improving forward radiative transfer calculations occurred. The
development of these techniques was driven, among others, by
\citet{1988ApJ...324..441S,1990MmSAI..61..559S,1996mpsa.conf..367S,1988ApJ...330..274B,1994ASPC...64..438L,1997A&A...318..429R,2000ASPC..198..371J}.

The displacement of a circularly polarized $\sigma$-component from line 
center can be expressed by:
\begin{equation}
  \Delta \lambda_{B} \approx 4.67 \times
  10^{-13}\,g_{\rm{eff}}\,B\,\lambda^{2}\,\,\,[\AA].
  \label{eq:ZeemanBroadening}
\end{equation}
For values typical of optical spectral lines, Eq. 
\ref{eq:ZeemanBroadening} demonstrates that even a comparatively 
strong field field along LOS is hard to detect. $Bf = 1$\,kG would 
produce a Zeeman splitting of merely  $\lambda_{B} = 17$\,m\AA\,for 
the $\sigma$-components of
a normal Zeeman triplet of a hypothetical line at $6000$\,\AA\,with
$g_{\rm{eff}} = 1.00$, or $\lambda_{B} = 42$\,m\AA\,in the case of a 
more sensitive $g_{\rm{eff}} = 2.50$ line. This should be compared to 
the line width of a typical stellar line profile broadened
thermally and due to turbulence by 2-6\,km\,s$^{-1}$ (i.e. 30-60\,m\AA), as 
well as by rotation between 0 and 5\,km\,s$^{-1}$ (i.e. up to 
100\,m\AA). In addition, the resolution
element at even very high resolving power, $R$, is comparable in size, 
e.g. $\delta\lambda = 27$\,m\AA\,for $R = 220\,000$.

The present paper has a two-fold purpose. On the one hand, it is in the
tradition of increasing the realism and accuracy of magnetic field measurements
on cool stars. It applies state
of the art inversion techniques to determine the magnetic field. Also,
from observations of the Sun it is clear that magnetic fields are associated
with significant brightenings or darkenings, which are generally neglected
when measuring stellar magnetic fields. In the present paper we have carried
out first simple computations looking at the effects that including such a
diagnostic would have. We do not propose that this is an exhaustive or final
study, but it can be considered as a first indicator.
Finally,  these techniques are also applied to spectral time series of an
active star that has so far not been studied in this respect.


\section{Data}\label{sec:two}
\subsection{Instruments and reduction}

Spectra of the stars investigated were obtained with the Coud\'e \'Echelle 
Spectrometer (CES) at the $3.6$\,m ESO
telescope at La Silla, Chile. This fiber-fed spectrometer achieved a
spectral resolving power of approximately $220\,000$ using an
image-slicer and projecting only part of an order onto the CCD,
resulting in a small wavelength coverage of approx.\,$40$\,\AA.

\begin{figure*}
  \includegraphics[clip]{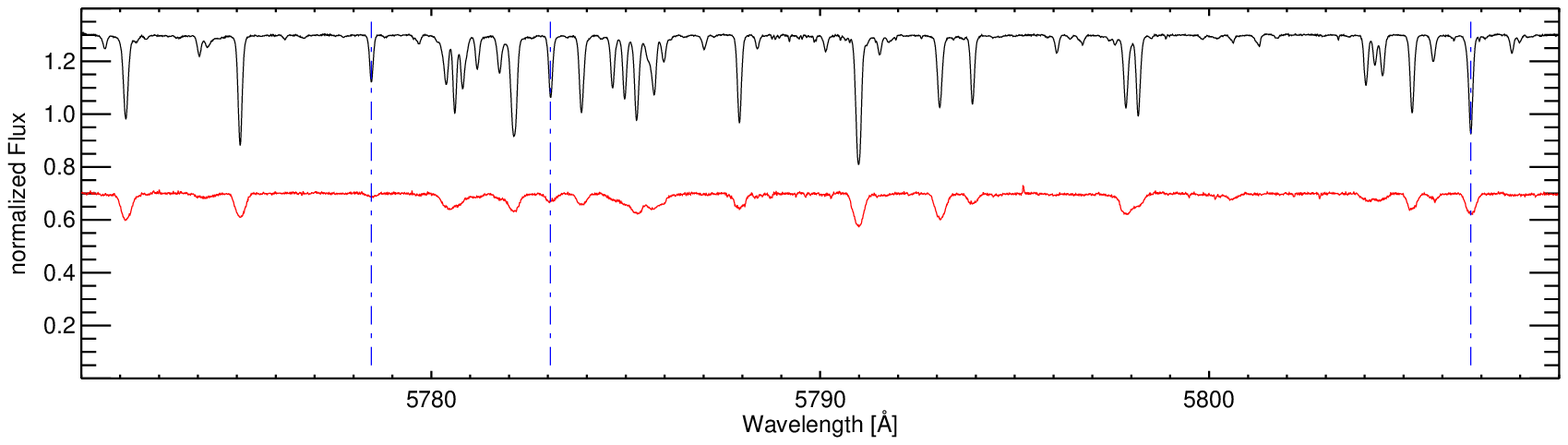}
  \includegraphics[clip]{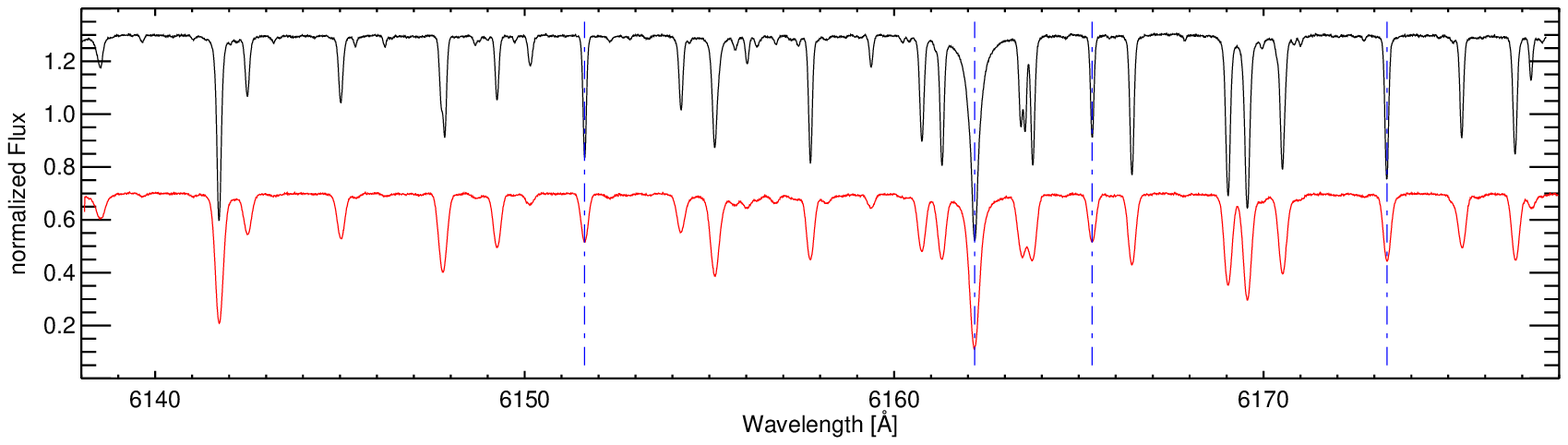}
  \caption{Sample CES spectra illustrating data quality. Continua are
    shifted by $\pm 0.3$ from unity for visibility. Blue dash-dotted
    vertical lines indicate spectral lines used for
    analysis.
    \textit{Upper panel:} data set A. Upper black spectrum: sunlight
    reflected from Jupiter's moon Ganymede; lower red spectrum:
    HD\,68456 (F6V, $v\sin{i}\approx 10$\,km s$^{-1}$).
    \textit{Lower panel:} data set
    B. Upper black spectrum: inactive and slow rotator 61\,Vir (G6V);
    lower red spectrum: active and faster rotator 59\,Vir (G0V,
    $v\sin{i} \approx 6.5$\,km\,s$^{-1}$).}
  \label{fig:SampleSpec}
\end{figure*}

Spectra from two data sets were used in this work: data set A, with
$5770 < \lambda < 5810$\,\AA, observed on 13 October 2000
(solar spectrum) and 2 October 2001 (HD\,$68456$), and data set B,
$6137 < \lambda < 6177$\,\AA,
observed on 7 May 2005 (HD\,$68456$, $59$\,Vir). Sample
spectra from both sets are shown in Fig. \ref{fig:SampleSpec}.  We
refer to \citet[hereafter R\&S03]{2003A&A...398..647R} for more
details, as data set A was originally taken for the project described
in that publication.

Where necessary, continuum normalization corrections were performed by
division of linear slopes fitted to the original continuum. Spectra
were shifted to laboratory rest-wavelengths, by cross-correlation
using the FTS solar atlas \citep{SolarFTS}.

\subsection{Objects, spectral lines and atomic line information}
In this work, we present results for 4 objects ranging in spectral
type from F6V to G6V, including sunlight reflected from Ganymede
(previously and hereafter referred to as the solar spectrum). These stars were
especially suitable for our analysis for multiple reasons. First,
identification of spectral line blends was facilitated in the
effective temperature range covered by these stars and LTE atmospheres
were readily available \citep{1992IAUS..149..225K}. Second, the
analysis of stars with relatively low projected rotational velocities,
$v\sin{i}$ $\lesssim 10$ km s$^{-1}$, allowed us to fully exploit the
high resolving power of the data and minimizes effects 
due to differential rotation. Third, measured stellar parameters such
as effective temperature, $T_{\rm{eff}}$, surface gravity, $\log{g}$
and elemental abundances, [Fe/H] in particular, were available as
constraints and could be directly input into the inversion procedure.
Last, but not least, measured X-ray
luminosities, $L_{X}$, and previously reported magnetic field
detections could be considered as indicators for the presence of magnetic
fields \citep{1990MmSAI..61..559S} and thus helped to constrain the
sample.

\begin{table}[]
  \begin{center}
    \begin{tabular}{lcccccc}\hline \hline \noalign{\smallskip}
      Line & $\lambda$ & Ion & $\log{(g^*f^*)}$  & $\chi_{e}$ & $g_{\rm{eff}}$ \\
      Set & $[$\AA$]$   &   &  & $[$eV$]$  \\
      \noalign{\smallskip} \hline \noalign{\smallskip}
      A3: & $  5778.46   $ &  Fe I      & $ -3.53  $ & $ 2.59 $ & $ 1.21 $ \\
      \noalign{\smallskip}
      &$  5783.06   $ &  Cr I      & $ -0.57 \,(-0.50) $ & $ 3.32 $ & $2.00$\\
      &$  5783.11   $ &  Cr I      & $ -1.32 \,(-1.82) $ & $3.56$ & $0.75$\\
      \noalign{\smallskip}
      &$  5806.73   $ &  Fe I      & $ -0.99 $ & $ 4.61 $ & $ 0.92 $ \\
      &$  5806.63   $ &  Fe I      & $ -2.19 \,(-2.09)$ & $4.91$ & $1.50$\\
      &$  5806.73   $ &  Sc II     & $ -3.02 $ & $1.36$ & $1.67$ \\
      \noalign{\smallskip} \hline \hline \noalign{\smallskip}
      B3: &$  6151.62   $ &  Fe I      & $ -3.30 $ & $ 2.18 $ & $ 1.83 $  \\
      \noalign{\smallskip}
      &$  6165.36   $ &  Fe I      & $ -1.51  $ & $ 4.14$ & $ 1.00 $ \\
      &$  6165.12   $ &  Fe I      & $ -2.76 $ & $5.09$ & $0.67$ \\
      \noalign{\smallskip}
      &$  6173.34   $ &  Fe I      & $ -2.88 $ & $ 2.22 $ & $ 2.5 $  \\
      &$  6173.03   $ & Eu II     & $ -0.86  $ & $1.32$ & $1.70$ \\
      \noalign{\smallskip} \hline \noalign{\smallskip}
      B4:& \multicolumn{4}{l}{Same as B3 plus}\\
      \noalign{\smallskip}
      &$  6162.18   $ &  Ca I      & $ -0.17 $ & $ 1.90 $ & $ 1.25$  \\
      \noalign{\smallskip}
      \hline \hline
    \end{tabular}
  \end{center}
  \caption{
    Spectral lines and their parameters used in inversions. Horizontal
    lines group line sets, double horizontal lines further
    distinguish the two data sets. Spectral lines are separated by
    small spacings indicating which blends belong together. Dominant
    line components
    are listed first, blends follow by ascending wavelength.
    Brackets indicate the closest $\log{(g^*f^*)}$ (oscillator strength)
    literature value, if values outside the literature range were adopted. }
  \label{tab:atomic}
\end{table}

Within the small wavelength coverage of the CES spectra, only few
relatively blend-free lines could be identified. Among these, we
searched for combinations of lines that provide strong constraints on
atmospheric properties. For instance, lines of different excitation
potentials, $\chi_{e}$, have different temperature sensitivities and
often probe different heights (depending also on their equivalent widths).
Therefore they constrain temperature stratification. In order to 
exploit the differential effect of Zeeman broadening, we aimed to cover the
largest possible range in $g_{\rm{eff}}$ as well as to use the line with 
the greatest possible $g_{\rm{eff}}$.

Most spectral lines could be identified in the postscript version of the FTS 
solar atlas available online. The remaining lines and blends were 
identified by searching the Vienna \citep[][hereafter VALD]{VALD} 
and \citet{KALD} atomic line databases. 
These databases further provided the atomic line data, such as level 
configurations and $\log{(g^*f^*)}$ values
required for the forward radiative transfer calculations, cf.
Sect.\,\ref{sec:three}. Effective Land\'e values, $g_{\rm{eff}}$, were
calculated from the atomic level configurations following
\citet{Beckers:1969}. The calculated effective and empirical Land\'e
factors determined by \citet{1985A&A...148..123S} were found to agree
for all but one line: Fe\,I at $6165.36$\,\AA, for which
$g_{\rm{emp}}=0.69$ and $g_{\rm{eff}}=1.00$. However, this line
contains a blend at $6165.36$\,\AA, for which $g_{\rm{eff}}=0.67$, cf.
Table\,\ref{tab:atomic}.

For some lines, ranges of $\log{(g^*f^*)}$ values were found in
\texttt{VALD}. We aimed to determine the best $\log{(g^*f^*)}$ value for our
selected lines by means of a by-hand gauging procedure using the solar
spectrum for data set A and that of 61\,Vir for data set B, while
taking care to stay as close to the literature values as possible.
This approach was adopted in order to avoid multiple additional free
fit parameters, since one $\log{(g^*f^*)}$ for each blend would have been
required. An automated approach would have been dominated by
degeneracy between oscillator strengths and stellar model parameters.

Table\,\ref{tab:atomic} lists all lines used and parameters adopted in
our inversions.
Horizontal spacings help distinguish between blends contributing to
the shape of a given spectral line and different lines. Col.\,1
lists abbreviations used in this paper for the respective line
sets. Col.\,2 contains wavelength in \AA, Col.\,3 the corresponding ion
in spectroscopic notation, i.e. Fe\,I is
neutral iron. Col.\,4 lists $\log{(g^*f^*)}$ adopted after gauging with the
closest literature value in brackets, if a value outside the
\texttt{VALD} range was adopted. Cols.\,5 and 6 contain
excitation potential of the lower level of the transition, $\chi_{e}$,
and effective Land\'e factor, $g_{\rm{eff}}$, respectively. The range
in Land\'e factor covered by
line set A3 is thus $\Delta g_{\rm{eff}} = 1.08$, and $\Delta
g_{\rm{eff}} = 1.50$ for line sets B3 and B4, not counting blends. More
importantly, the maximum $g_{\rm{eff}}$ is larger by 0.5 in line sets B3 and
B4 than in line set A3.
Therefore, a stronger sensitivity to ZB in the latter
two data sets can be expected.

Collisional damping parameters are derived internally by
\texttt{SPINOR} from cross-sections published by
\citet{1995MNRAS.276..859A,1997MNRAS.290..102B,1998MNRAS.296.1057B}
and are not listed in the table.

\begin{table}[]
  $$
  \begin{array}{llccccc}
    \hline\hline
    \noalign{\smallskip}
    \rm{Name} & \rm{HD} & \rm{[Ca/H]} & \rm{[Sc/H]} & \rm{[Cr/H]} &
    \rm{[Fe/H]} & \rm{[Eu/H]} \\
    \noalign{\smallskip}
    \hline
    \noalign{\smallskip}
    \rm{Sun} & - & 6.36 & 3.17 & 5.67 & 7.50 & 0.51 \\
    \noalign{\smallskip}
    \hline
    \noalign{\smallskip}
    \rm{GJ}~297.1 & 68456 & 6.10 & 2.90 & 5.37 & 7.19 & 0.35 \\
    \noalign{\smallskip}
    \hline
    \noalign{\smallskip}
    59~\rm{Vir} & 115383 & 6.57 & - & - & 7.78 & 0.70 \\
    \noalign{\smallskip}
    \hline
    \noalign{\smallskip}
    61~\rm{Vir} & 115617 & 6.36 & - & - & 7.55 & 0.50 \\
    \noalign{\smallskip}
    \hline
    \hline
  \end{array}
  $$
  \caption[]{Logarithmic elemental abundances adopted on scale where
  hydrogen equals 12.}
  \label{tab:abunds}
\end{table}

Table\,\ref{tab:abunds} lists the elemental abundances adopted. Solar
abundances used were those determined by
\citet{1998SSRv...85..161G} and served as the absolute scale for
relative stellar literature values found in \citet[from hereon
V\&F05]{2005ApJS..159..141V} (59\,Vir \& 61\,Vir) and
\citet{1997A&AS..124..299C} (HD\,68456).

Lines that correspond to elements for which no abundance measurements were
found were usually not included in the inversions. For the Cr\,I line in A3
as well as some blends, e.g. Eu\,II in B3, abundances were estimated from an
iterative fitting procedure that took into account published average
metallicities. This procedure is potentially flawed, since general abundances
are not necessarily correlated with individual abundances. However, this flaw
does not challenge the work presented here, since Fe\,I lines were the most
important ones for our analysis and literature values for iron were available
for all stars. The Cr\,I line was of great importance only for the solar case,
for which measurements of the chromium abundance (and also that of scandium)
do exist. The blend due to Eu\,II was only of secondary significance, given
its small impact on line shape. It was not discarded, however, since it
provided continuum points for (at least the blue wing of) the Zeeman-sensitive
$6173.3$\,\AA\ line, for which the red wing was cut off due to the presence of
unidentified blends.

One well known Zeeman sensitive line was
  omitted from the analysis, Fe\,II at $6149.24$\,\AA. This line is
  very sensitive to magnetic fields due to its particular (doublet)
  Zeeman pattern, and displays 
  clear Zeeman splitting in spectra of Ap stars
  \citep[e.g.][]{1990A&A...232..151M,1997A&AS..123..353M} and  
solar penumbrae \citep[e.g.][]{1991ApJ...373..683L}. Unfortunately, 
this line is affected by blends in both line wings, as can be seen in
Fig.\,\ref{fig:SampleSpec} and we were not able to adequately
reproduce these blended wings. While the blue wing blend is quite
obvious, the red wing blend requires closer inspection. This blend is
(at least in part) due to Ti\,I at $6149.73$\,\AA\,and is hard to see due to 
rotational broadening, blending in with the smaller
unidentified signatures seen in the slower rotator $61$\,Vir. This
leads to the wing of the line just barely missing the
continuum. Due to these two blended wings, the $6149.24$\,\AA\,Fe\,II
line profile used for inversion had either 
a badly defined continuum (the deviation from the continuum is
around $1\%$), or lacked both line wings (if they were both cut off as
was done for the red wing of the $6173$\,\AA\,Fe\,II line) which meant
not having any continuum points at all. Using either approach, we
were  not able to reach the same level of fit precision as presented
for the other lines used in this analysis. Therefore, we chose to omit
this line from our analysis.

Our results depend on the assumed Signal-to-Noise ratio (S/N), since
\label{sec:sndep}
(S/N)$^{-1} = \sigma_{i}$ and
\begin{equation}
\chi^{2} = \sum_{i,j} \frac{\left(I_{ij} - I^{'}_{ij}\right)^{2}}{\sigma_{i}^2},
\label{eq:chi2}
\end{equation}
where $\sigma_{i}$ is significance, $i$ line index, $j$ data point
index for a given line, $I_{ij}$  measured  and
$I'_{ij}$ calculated intensity normalized to the continuum value.
We determined $\sigma_{i}$ from the
mean continuum standard deviation of the spectra. The typical S/N ratio 
thus determined was $400 - 700$, corresponding to $\sigma$
between $0.0025 - 0.00143$. 
Less statistical weight was assigned to the pressure-sensitive Ca\,I
line at $6162.18$\,\AA\,by lowering significance to $\sigma = 0.0040$,
since the wings of the line are heavily blended and asymmetric.
Furthermore, the core of this line is affected by non-LTE effects and
was therefore removed from the measured line profile.

In every line, we aimed to take full account of line blends and
included as many data points reaching as far as possible into the
continuum, since ZB affects the far line wings most
strongly. However, some line wings had to be severed farther from
continuum than others, since unidentified blends were present; see, for
instance, the far red wing of the Zeeman-sensitive line at
$6173.34$\,\AA~Fe\,I.  This line wing showed strong contributions
from at least one blend that could not be identified in line
databases. To circumvent this problem, \citet{1997A&A...318..429R}
included fake iron blends until the line shape was matched. We chose not
to introduce additional fit parameters, and instead decided to sever the
far red line wing.

\section{Analysis}\label{sec:three}
We use spectral line inversion as the tool to explore detectability of
ZB in this paper. Spectral line inversion denotes
a combination of two things: forward radiative transfer calculations
and an iterative non-linear least squares fitting process. In other
words, spectral line profiles are calculated using a given atmospheric
model and model parameters are iteratively adjusted until the highest
possible agreement between calculated and observed spectral lines is
reached. The resulting best-fit model parameters thus characterize the
adopted atmospheric model and, if the model is assumed to be correct,
the actual stellar atmosphere. Hence, we investigate whether introducing a
magnetic field significantly improves the agreement between observed and
calculated line profiles. We explain the general approach of spectral line
inversion in the following subsection.

We further investigate model-dependence and the influence of model
assumptions. To this end, two different kinds of models are
used. These fitting strategies are documented in
Sect.\,\ref{sec:FittingStrategies}.

\subsection{Spectral line inversions}
We employ the spectral line inversion package \texttt{SPINOR}
developed and maintained at ETH Z\"urich and MPS Katlenburg-Lindau for
this analysis. In short, \texttt{SPINOR} computes
  synthetic intensity profiles (SIPs), $L(\lambda)$, by solving the
  radiative transfer equation on a grid of optical depth 
points $\tau$, accounting also for turbulence, rotation, magnetic
fields, instrumental effects, etc. It then determines the model
parameters used to compute $L(\lambda)$ that best fit the observed
data using a non-linear least squares fitting algorithm.
We now briefly discuss how the SIPs used for inversion are
calculated and refer to \citet[from hereon Fru00 and Fru05,
  respectively]{2000A&A...358.1109F,2005A&A...444..549F} for further
information regarding \texttt{SPINOR}.

First, annular SIPs, $I(\lambda,\theta)$, are computed as a sum of 
weighted Voigt profiles \citep[cf.][from hereon
  Gr08]{2008oasp.book.....G} combining thermal and pressure  
broadening, microturbulence (mass motions on a scale that is much
smaller than the optical depth lead to a Doppler broadening due to a
Gaussian velocity dispersion),  $v_{\rm mic}$, and the full Zeeman 
pattern as described below. ``Annular'' here denotes the
$\theta$-dependence, since synthetic profiles are calculated for each
of $15$ concentric annuli that taken together approximate a flux
profile resulting from a much larger number of intensity profiles at
different viewing angles \citep[Fru05]{1997A&A...318..429R}. By
calculating annular SIPs at different angular distances from 
disk center, we can include limb darkening due to its dependence on
$\mu = \cos\theta$. 

To calculate the radiation transfer, we start with tabulated LTE
  atmospheres \citep{1992IAUS..149..225K}. These
  atmospheres list the parameters temperature, pressure and electron
  pressure for a number of $\tau$ values and are calculated for given
  effective temperatures, $T_{\rm{eff}}$, and continuum opacity at 
$5000$\,\AA. In general, the temperature values are allowed to vary in
  order to obtain an optimal fit to the spectral lines. Then, the
  whole temperature stratification $T(\tau)$ is shifted up or
  down. At each step in the inversion the atmosphere is always
  consistently maintained in hydrostatic equilibrium. Hence, from a
  prescribed temperature stratification, the gas and
electronic pressures, the gas density and continuum opacity are
  computed assuming hydrostatic equilibrium.
Note that temperature values returned by \texttt{SPINOR}
correspond to unit optical depth calculated for continuum opacity at
$5000\,$\AA. 

The Zeeman shifts and amplitudes of the polarized components
  are calculated as explained in, e.g. \citet{1974SoPh...35...11W}.
A detailed
description of the implementation of the Zeeman effect in
\texttt{SPINOR} can be found in \citet{Frutithesis} and references
therein. 

In a next step, the individual annular $I(\lambda,\theta)$ are
convolved with radial-tangential macroturbulence profiles,
$\Theta(\lambda,\theta)$, that are intended to account for 
large-scale motion due to convection. These profiles are described in  
\citet{1975ApJ...202..148G} and Gr08. Thus, we obtain new annular SIPs,
$I'(\lambda,\theta) = I(\lambda,\theta)\, *\,
\Theta(\lambda,\theta)$, where ``$*$'' denotes convolution carried
out as multiplication in Fourier space. In accordance with Fru05, we
assume equal surface fractions covered by radial and tangential flows.

The radial velocities of individual surface elements vary across the
stellar disk as a function of distance to the rotation axis, leading to
rotational broadening in the integrated spectrum of a point source. This
dependence is taken into account by introducing $15$ new angles, $\phi$,
so that the radial velocity is constant at any given $\phi$. Thus, the
$I'(\lambda,\theta)$ are turned into $I'(\lambda,\theta,\phi)$ and each
one is shifted for radial velocity. The disk-integrated line profiles
are obtained by weighting the individual $I'(\lambda,\theta,\phi)$ for
surface coverage, and summing up. 

We obtain the final SIP, $L(\lambda)$, by computing $L(\lambda) =
 I'(\lambda)\, *\, I_{\rm{inst}}$, that is by convolution of $
 I'(\lambda)$  with an instrumental broadening function,
 $I_{\rm{inst}}$. $I_{\rm{inst}}$ is a Gaussian function with a FWHM
 defined as a velocity, $v_{\rm{inst}}$, in km s$^{-1}$ that
 corresponds to the resolving power of CES. Hence, we have $R =
 220\,000 = \lambda / \Delta\lambda = c / v_{\rm{inst}}$. Therefore,
 for CES, $v_{\rm{inst}} = 1.36\,\rm{km\,s}^{-1}$. 

\label{sec:fitparms}
Summing up, our model parameters are: temperature $T_1$ (and
  $T_2$ for two-component inversions, cf. Sect. \ref{sec:twocomp}),
  in Kelvin at unit optical depth; 
surface gravity $\log{g}$ in cgs units; projected rotational
($v\sin{i}$), microturbulent ($v_{\rm{mic}}$) and 
macroturbulent ($v_{\rm{mac}}$) velocities, each in km s$^{-1}$.
$\log{g}$ was kept fixed at literature values in most inversions,
since narrow Fe\,I lines do not provide strong constraints on
surface gravity. Inversions of line set B4, however, allowed
$\log{g}$ to vary, since the pressure-sensitive Ca\,I line did provide
this constraint.  Finally, lines were allowed to be shifted in
velocity space by $\leq 500$\,m\,s$^{-1}$ $(\sim 0.01\AA {\rm\ at\ } 6000\AA)$
to correct for slight differences in literature and data line center
wavelengths, as well as possible shortcomings in the dispersion
relation.

Our treatment is state-of-the-art for a line
inversion technique and motivated by computational efficiency without
departing from physical realism. However, it has to be kept in mind
that the line profiles described here  
are approximations that rely on a particular implementation
(e.g. of the radial-tangential macroturbulence, or the
height-independence of other parameters). This will be a limiting
factor for the achievable fit precision and we can expect to find
some residual signatures of this treatment in our analysis. Thus, an
important part of interpreting our results will be linked to 
distinguishing these effects from a ZB signature.

\subsection{Fitting strategies and atmospheric models}
\label{sec:FittingStrategies}
Given the highly
complex field topologies expected for sun-like stars, most past
techniques to measure ZB assumed that coverage by
magnetic regions was homogeneous and atmospheres identical in all but
their magnetic properties. Usually, both magnetic field
strength, $B$, and surface area covered by magnetic regions (filling
factor), $f$, were then determined independently, while the effects
of multi-modal surface temperature distributions were neglected. These
models computed the total radiative flux from the star as the sum of flux from
magnetic and non-magnetic (quiet) surface fractions with identical
temperature stratification:
\begin{equation}
F_{\rm{tot}} = (1-f)F_{\rm{quiet}} + fF_{\rm{magn.}}.
\label{eq:ffac}
\end{equation}

To investigate detectability, we chose two different 
strategies. First, models with a single homogeneous atmosphere were
considered, so called one-component (OC) models, where only the
product $Bf$ was determined. Second, two-component (TC) models were used, where
$B$ and $f$ are naturally segregated, since $f$ defines the relative
surface fractions covered by the respective components, cf. Eq.\,\ref{eq:ffac}.
A further refinement of this model takes into account the generally different
temperature stratifications of magnetic and non-magnetic regions. This
temperature difference is well-known from the Sun (see, e.g., Solanki 1993 for
an overview). This
dichotomy in fitting strategies is addressed in the following and
results are presented in separate subsections.

For both strategies adopted, we allow all fit parameters aside from
$Bf$ (OC) or $B$ and $f$ (TC) to vary freely in
the fit (see last paragraph of Sect.\,\ref{sec:fitparms}). However, as
explained above, $\log{g}$ was frequently used at a fixed
literature value. Thus, we calculate $\chi^{2}$, cf.
Eq.\,\ref{eq:chi2}, for a range of values of $Bf$ that we keep fixed
in each inversion. In this way, we aim to avoid degeneracies that
prevent the non-linear least-squares algorithm from finding the global
minimum. This strategy implies that we do not aim to determine the
other model parameters accurately. Our goal is to investigate whether
introduction of magnetic flux (and how much) in the atmospheric model
yields significantly better results.

To quantify our results, we determine the overall
best-fit solution, i.e. where $\chi^{2}$ reaches its global minimum,
as well as formal errors on $Bf$ following \cite{1992nrca.book.....P}.
The formal $1\sigma$ range is adopted as our standard error. To claim
detection of magnetic fields, we require the non-magnetic
best-fit solution to lie outside the $3\sigma$ confidence
level (CL). Non-detections are characterized by $3\sigma$ upper limits.

The CLs on  $\chi^{2}$ are of a formal nature,
since they depend on $\chi^{2}_{\rm{min}}$, cf. Sect.\,\ref{sec:sndep}.
Errors deduced from these CLs would therefore represent the
true errors only in the case of a ``perfect'' fit. In addition, the choice
of a specific (Kurucz) input atmosphere
introduces a bias that affects $\chi^2$ and thereby
influences the formal error determination on $Bf$.
However, spectral lines differ in their respective sensitivities 
to magnetic fields (usually expressed by 
their effective Land\'e factor, cf. Table\,\ref{tab:atomic}). We employ
both Zeeman sensitive and insensitive lines in simultaneous
inversions. Each line's different reaction to a certain
postulated magnetic flux directly impacts the obtained best-fit
$\chi^2$ for this $Bf$ value. Therefore, provided the inversions are
not dominated by systematic effects (see, however,
Sect.\,\ref{sec:SunTC}), the
CLs do allow us to evaluate whether evidence for a magnetic field is
found, despite the above statements about their mostly formal nature. The
CLs further provide an order-of-magnitude estimate on the uncertainty
involved in the measurement. 

\subsubsection{One-component inversions}\label{sec:onecomp}
In the OC approach, we scan the dependence of $\chi^{2}$ on the single
parameter $Bf$. This is done by performing inversions for a set
of fixed $Bf$ values. Best-fit solutions for $\chi^{2}$ and model
parameters are obtained from each inversion.

In OC inversions we formally set $f = 1$ and investigate the line
shaping effect due to the product $Bf$. Therefore, we assume a mean
magnetic field covering the entire surface, aiming to avoid the
degeneracy between $B$ and $f$. The spacing of the $Bf$ grid is
usually $50$\,G, and is refined
to $25$\,G close to the overall best-fit $\chi^{2}_{\rm{min}}$ for
better sampling of $\chi^{2}$ in this region.

\subsubsection{Two-component inversions}\label{sec:twocomp}
Following the solar paradigm, magnetic fields in cool stars are expected to
be concentrated in distinct elements that are usually described as flux tubes
\citep{1973SoPh...32...41S, 1976SoPh...50..269S, 1993SSRv...63....1S}. 
Our two-component (TC) inversions aim to take into
account this physical fact by composing the atmospheric model of a magnetic
and a field-free component. Thereby, we can investigate the
impact of the $f=1$ assumption made for OC models as well as the effect of
a bimodal surface temperature distribution.

Hence, we perform three cases of TC inversions: those where
both components have identical temperature, as well as
those where magnetic regions are either cool, or warm
with respect to the non-magnetic component. The equal
temperature case was frequently used in the literature and allows for a more
direct comparison of our results than OC inversions. Our 
incremental approach of increasing model complexity aims to distinguish between
geometrical and temperature-related effects.

\begin{figure*}[]
  \centering
  \includegraphics[clip]{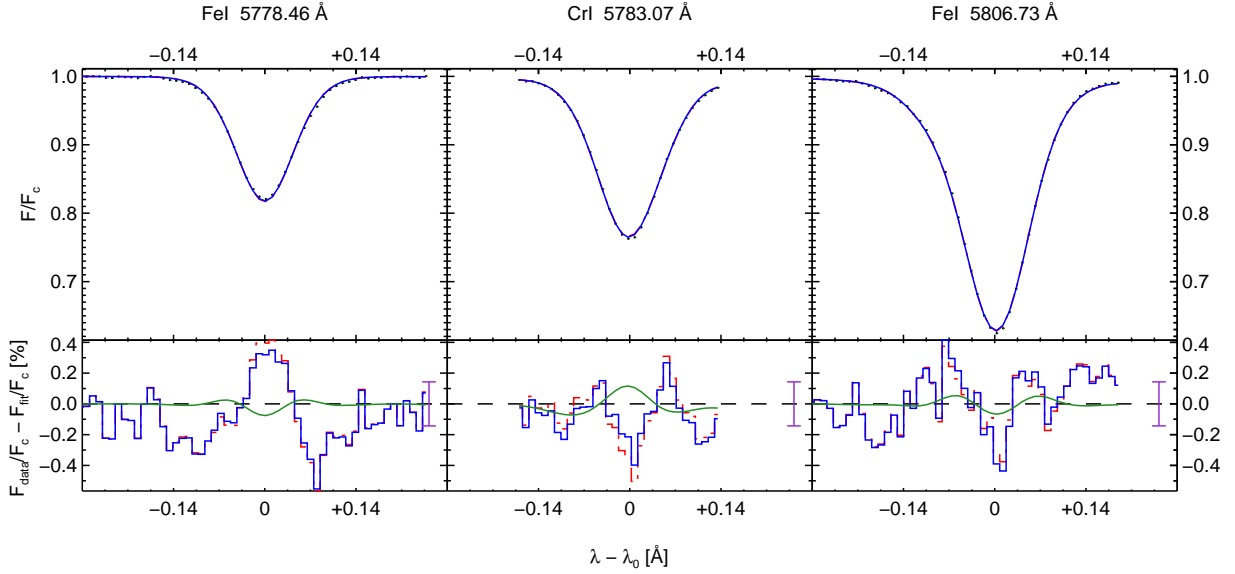}
  \caption{Sun, line set A3, one-component model: Data and
    best-fits with $Bf = 0$\,G (solid blue) and $Bf = 225$\,G
    (dash-dotted red)} \label{fig:GanyA3}
\end{figure*}

We compute the best-fit  $\chi^{2}$ for each point
of the $B$-$f$-plane defined by a grid of fixed $B$ and $f$ values.
As a result, this approach is computationally much more expensive.

In the TC inversions, the model contains surface filling
factor, $f$, as an additional fixed parameter, since we scan the
$B$-$f$-plane. In addition to $f$, we introduce the additional fit
parameter $T_2$ for temperature at unit optical depth for the second
component. For the equal temperature  
case, $T_2$ and $T_1$ are fitted, but kept equal. Both parameters can vary
freely and independently in the cool and warm cases.
To quantify our results, we construct formal CLs
on the two-dimensional $\chi^{2}$-maps in analogy to the OC inversions.

In this paper, we perform TC inversions for the Sun and
59\,Vir. For 59\,Vir, for which a magnetic field is detected with
a high CL, we investigate and compare the three different
TC cases discussed above.


\section{Results}
We present our results in two kinds of figures, namely line fit plots
and $\chi^{2}$-plots or -maps. Figure\,\ref{fig:GanyA3} is an example of
line fit plots that help to assess the qualitative difference
between magnetic and non-magnetic best-fit models. The upper panels
contain the measured line profiles as ticks of length $\sigma_{i}$
overplotted by best-fit solutions. Solid blue represents the overall
best-fit solution, and dash-dotted red lines are used to draw either the
non-magnetic, i.e. $0$\,G (if evidence for magnetic fields is found),
or formal $3\sigma$ upper limit on $Bf$. Residuals are drawn in
histogram style in the lower panels using the identical color and
line style theme and visualize the difference between measured and
calculated line profiles, scaled by factor 100. The purple error bar
at the right end in each of the residual panels indicates $\sigma_{i}$.

The residuals contain smooth solid green
lines that indicate the difference in line shape between the overall
best-fit and the other best-fit model plotted, i.e. the change in line
shape due to introduction of magnetic flux. The other fit parameters
such as $v\sin{i}$, etc., adapt to each fixed $Bf$ value, since
they are allowed to vary freely. Hence, the green line does not
represent the Zeeman pattern alone. It illustrates the combined effect
due to ZB and different best-fit parameters for the
two cases plotted.

Our inversions converged after relatively few iterations and although
a given choice of starting values for the fit parameters (especially
temperature) introduces a bias (if the $\chi^2$ surface has multiple
minima), no dependence on these was found for the best-fit parameters.

\begin{table*}[]
  $$
  \begin{array}{llp{0.04\linewidth}ccccccclcccccc}
    \hline\hline
    \noalign{\smallskip}
    \mathrm{Name} & \mathrm{HD} & Sp. & \mathrm{line}  &
    \chi^{2}_{\mathrm{min}} & N_{\rm{DOF}} & \rm{S/N} & T_{\rm{eff}} &
    T_{1} & T_{2} &\log g & v_{mac} & v_{mic} & v \sin{i} &
    Bf_{\rm{lit}} & Bf & Bf (3\sigma)\\
    &  & Type & \mathrm{set} & &  &  &
    \multicolumn{3}{c}{\mathrm{[K]}} & \mathrm{[cgs]} &
    \multicolumn{3}{c}{\mathrm{[km\,s}^{-1}]} &  \multicolumn{3}{c}{\mathrm{[G]}}\\
    \noalign{\smallskip}
    \hline
    \noalign{\smallskip}
    \mathrm{Sun} & - & G2V & \mathrm{A3} & 237.6 &
    132 & 700 & 5780 & 6565 & & 4.44 & 3.29 & 0.65 & 1.34 &
    \lesssim 130^{\mathrm{a}} & 0^{+90} & 0 - 200 \\
    \noalign{\smallskip}
    \multicolumn{3}{r}{\rm{TC, cool}} & \mathrm{A3} & 169.5
    & 130 & 700 & & 6571 & 4876 & 4.44 & 2.80 & 0.61 & 1.93 & &
    245^{+5}_{-10} & 150 - 1400\\
    \noalign{\smallskip}
    \hline
    \noalign{\smallskip}
    \mathrm{GJ}~297.1 & 68456 & F6V & \mathrm{A3} & 245.6 & 227
    & 400 & 6396 & 7178 & & 4.14 & 1.52 & 3.30 & 9.48 & - &
    1100^{+350}_{-550} & 0 - 1950 \\
    \noalign{\smallskip}
    & & & \mathrm{B3} & 318.6 & 248 & 420 &  & 7249 & & 4.14 & 2.33 &
    3.05 & 9.39 & - & 1050^{+100}_{-110} & 750 - 1300 \\
    \noalign{\smallskip}
    & & & \mathrm{B4} & 384.5 & 309 & 420 &  & 7209 & &
    4.40^{\mathrm{*}} & 4.63 & 1.84 & 9.35 & - & 950^{+110}_{-80} &
    600 - 1200\\
    \noalign{\smallskip}
    \hline
    \noalign{\smallskip}
    59 \mathrm{~Vir}  & 115383 & G0V & \mathrm{B3} & 262.5 & 223 & 420 &
    6234 & 7075 & & 4.60 & 4.89 & 1.21 & 6.67  &  &
    500^{+40}_{-60} & 300 - 600  \\
    \noalign{\smallskip}
    &        & & \mathrm{B4} & 401.9 & 332 & 420 &  &
    7077 & & 4.50^{\mathrm{*}} & 4.76 & 1.23 & 6.70 & & 525^{+30}_{-60} &
    350 - 650 \\
    \noalign{\smallskip}
    \multicolumn{3}{r}{\rm{TC, equal}} & \rm{B3} & 262.5 & 222 & 420 &
     & 7075 & 7075 & 4.60 &  4.88 & 1.21 & 6.68 & 190^{\mathrm{b}} &
    420^{+30}_{-265} & 100 - 550 \\
    \noalign{\smallskip}
    \multicolumn{3}{r}{\rm{TC, cool}} & \rm{B3} & 216.2 & 221 & 420 &
     & 7517 & 5978 & 4.60 &  4.29 & 1.42 & 6.72 & & 120^{+70}_{-120} & 0 - 300
    \\
    \noalign{\smallskip}
    \multicolumn{3}{r}{\rm{TC, warm}} & \rm{B3} & 226.1 & 221 & 420&
    & 6047 & 7440 & 4.60 & 4.39 & 1.33 & 6.70 & & 270^{+60}_{-60} & 0
    - 450\\
    \noalign{\smallskip}
    \hline
    \noalign{\smallskip}
    61 \mathrm{~Vir}  & 115617 & G6V & \mathrm{B3} & 198.1 &
    175 & 400 & 5571 & 6284 & & 4.47 & 3.09 & < 0.10 & 0.46 &
    600^{\mathrm{c}} & 0^{+80} & 0 - 150 \\
    \noalign{\smallskip}
    \hline\hline
  \end{array}
  $$
  \begin{list}{}{}
  \item[$^{\mathrm{a}}$] \citet{2004Natur.430..326T};
    $^{\mathrm{b}}$ \citet{1994ASPC...64..438L} ; $^{\mathrm{c}}$
    \citet{1984ApJ...277..640G}; $^{\mathrm{*}}$ $\log{g}$ used as
    free fit parameter
  \end{list}
  \caption[]{Objects investigated sorted by increasing HD
    number. Values for $T_{\rm{eff}}$ and $\log{g}$ from literature,
    unless otherwise indicated. Note that $T_{1}$ and $T_{2}$ are
    \texttt{SPINOR} best-fit temperatures corresponding to unit optical depth
    calculated for continuum opacity at $5000$\,\AA. Overall best-fit results as well as
    $1\sigma$ and $3\sigma$ ranges determined for $Bf$, see
    text. $3\sigma$ $Bf$ range rounded to 50\,G.}
    \label{tab:results}
\end{table*}

Table\,\ref{tab:results} lists our results for $Bf$ and overall
best-fit parameters, i.e. those at $\chi^{2}_{\rm{min}}$, for both OC and TC
inversions. The overall
best-fit values for $B$ and $f$ from TC inversions are found in the
text in the respective subsections. Col.\,1, 2
and 3 contain the star's name, HD number and spectral type. Spanning
these columns, we indicate if and for what kind of TC model results
were obtained. The line sets used, as introduced in
Table\,\ref{tab:atomic},   are listed in Col.\,4.
Col.\,5, 6 and 7 list the overall best-fit $\chi^{2}_{\rm{min}}$,
number of degrees of freedom $N_{\rm{DOF}}$, and continuum
signal-to-noise (S/N) measured in the data.  Cols.\,8, 9 and 10 provide
literature values for effective temperature, $T_{\rm{eff}}$, as well as
overall best-fit temperature at unit optical depth computed for continuum
opacity at $5000$\,\AA, $T_{1}$, and, in
the case of TC inversions, $T_{2}$, in Kelvin,
cf. Sect.\,\ref{sec:FittingStrategies}. Col.\,11
contains literature values for logarithmic gravitational acceleration
in cgs units, unless marked by an asterisk (*). In this case,
$\log{g}$ was used as a free fit parameter, constrained by the
pressure-sensitive Ca\,I line, and the overall best-fit value is
presented. Col.\,12, 13 and 14 list the overall best-fit parameters
obtained for macroturbulence $v_{\rm{mac}}$, microturbulence
$v_{\rm{mic}}$, and projected rotational velocity, $v\sin{i}$, each in
km s$^{-1}$.
Col.\,15 contains literature values for $Bf$, where
available. Col.\,16 lists our results for average magnetic field
with formal $1\sigma$ errors obtained from interpolation of the
$\chi^{2}$-plots or -maps. The last Col.\,17 quotes
the $3\sigma$ range of $Bf$ rounded to $50$\,G.

Additional figures for OC results obtained can be found in the online
appendix. These include behavior of best-fit parameters with $Bf$ as
well as line profile plots for all OC results listed in
Table\,\ref{tab:results} that are not included in the main article body.

\subsection{One-component inversions}\label{sec:res:oc}
\begin{figure*}[]
  \centering
  \includegraphics[width=1.\textwidth]{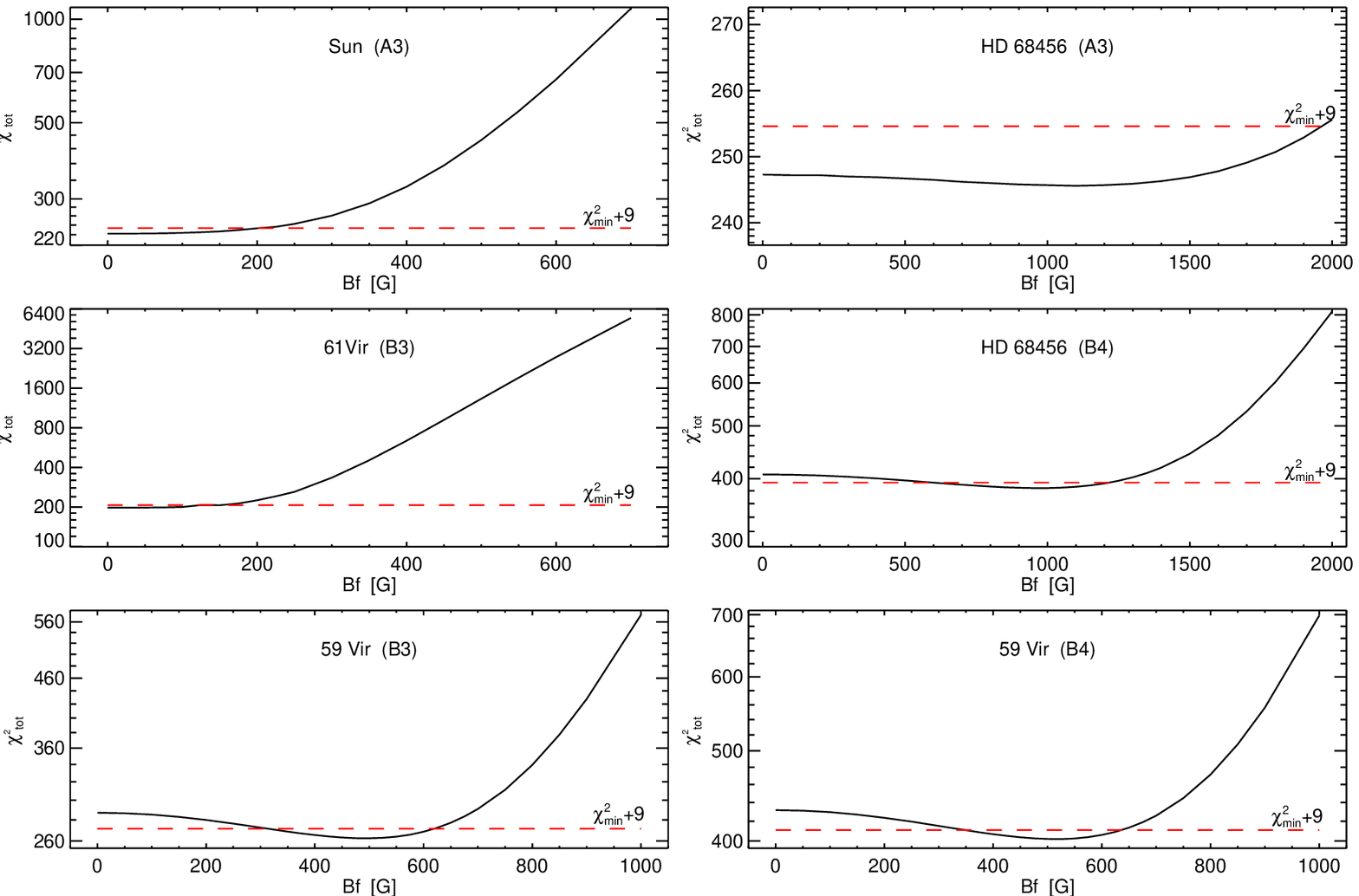}
  \caption[]{$\chi^{2}$-plots obtained from one-component
  inversions. Object names are given inside the respective panels, and
  line sets used are mentioned in parentheses. $3\sigma$ confidence
  level on $Bf$ indicated by dashed red horizontal lines.}
  \label{fig:chimaps}
\end{figure*}

OC inversions were carried out for 4 stars in different
line sets, see Table\,\ref{tab:atomic}. Figure\,\ref{fig:chimaps} shows
the $\chi^{2}$-plots determined for (from top left to bottom
right) the Sun (A3), HD\,68456 (A3), 61\,Vir (B3), HD\,68456 (B4),
59\,Vir (B3) and 59\,Vir (B4), where brackets indicate line sets
used. The dashed red horizontal lines indicate the formal $3\sigma$
CL defined by $\chi^{2}_{\rm{min}} + 9$, cf. Sect.\,\ref{sec:onecomp}.

\subsubsection{The Sun}\label{sec:Gany1comp}
The spectrum of sunlight reflected from Ganymede's surface is the
highest-quality spectrum used in this work with S/N $\sim 700$. We
adopted solar abundances as determined by \citet{1998SSRv...85..161G}
and \citet{1992IAUS..149..225K} input atmosphere \texttt{asun.dat13}.
Surface gravity was fixed at $\log{g} = 4.44$.

The upper left panel of Fig.\,\ref{fig:chimaps} clearly shows that the
non-magnetic model yields our overall best-fit result for the Sun. We
determine $Bf = 0^{+90}$\,G and a formal $3\sigma$ upper
limit: $Bf \leq 200$\,G. $\chi^{2}$ increases smoothly and steeply above
$200$\,G, indicating a robust non-detection. The overall best-fit
parameters are listed in Table\,\ref{tab:results} and are consistent
with the literature values published by V\&F05: $v\sin{i} = 1.63$\,km
s$^{-1}$, $v_{\rm{mac}} = 3.98$\,km s$^{-1}$, $v_{\rm{mic}} =
0.85$\,km s$^{-1}$.

Figure\,\ref{fig:GanyA3} shows the measured line profiles (black tick
marks) of line set A3 overplotted by non-magnetic (blue) and magnetic
(red, $Bf = 225$\,G) best-fit calculated profiles. The agreement
between either model and the measured line profiles is so close that
amplified residuals have to be considered to view the difference. The
most Zeeman-sensitive line, Cr\,I with $g_{\rm{eff}} =
2.00$ (center), exhibits the strongest reaction to
magnetic flux. This is indicated by the smooth green line. However,
the amplitude of this reaction is of the same order of magnitude as
$\sigma_{i}$.

A clear systematic signature is seen in the
residuals, notably in the $5778.46$\,\AA~Fe\,I line. This W-shaped
signature is not removed by ZB in this OC inversion and
is roughly twice as large as the solid green line that indicates the
difference between the non-magnetic and magnetic models. This
indicates a systematic shortcoming in our ability to reproduce line
shape and may indicate that, for instance, our approximate
treatment of convection (as a macroturbulence) may be inadequate at
this level of precision.

From this analysis, which is consistent with the upper limit of 140 G set by
\citet{1977A&A....59..367S}, also using broadening of intensity profiles,
we can neither confirm nor rule out the result by
\citet{2004Natur.430..326T} of $Bf\sim 130$\,G, since this value is
consistent with our formal $3\sigma$
CL. The same is true for other, generally smaller, literature values
of average (turbulent) magnetic fields determined for the
quiet Sun \citep{2009ASPC..405..135S}. We conclude that magnetism averaged
over the solar surface is too weak to be detected using this line set
and method. Note that the above literature values refer to the quiet Sun,
whereas the Sun was rather active at the time the analyzed data were
recorded (13 October 2000).  However, adding the
active-region magnetic flux would raise the $Bf$ value by a couple of $10$\,G
only and thus would not affect our conclusions.

\subsubsection{61 Vir -- HD 115617}
The spectrum of the G6V field dwarf 61\,Vir in data set B was an ideal
candidate for a robust magnetic field measurement due to its slow
rotation (R\&S03: $v\sin{i} <2.2$\,km s$^{-1}$), close-to solar
metallicity ([Fe/H] = 0.05), $T_{\rm{eff}} = 5571$\,K (latter two in
V\&F05), and low X-ray activity (NEXXUS2 database
\citep{2004A&A...417..651S}: $\log L_{\rm{X}} = 26.65$). We used an
input atmosphere calculated for $T_{\rm{eff}} = 5500$\,K.  $\log g = 4.47$
(V\&F05) was used as a fixed parameter.

The $\chi^{2}$-plot in the center left panel of
Fig.\,\ref{fig:chimaps} confirms the robust non-detection
for 61\,Vir. This is in agreement with expectations motivated 
by findings that link X-ray luminosity with magnetic flux 
\citep{2003ApJ...598.1387P}. The 
overall best-fit is reached for $Bf = 0$\,G and formal 
$1\sigma$ and $3\sigma$ upper limits
are established at $Bf = 80$ and $150$\,G, respectively. See
Table\,\ref{tab:results} for details on other best-fit parameters. The lower
$3\sigma$ upper limit obtained for 61 Vir than for the Sun can be explained
by the higher Zeeman sensitivity of the line set B3 (61 Vir) relative to A3
(Sun).

We found one previous measurement of magnetic flux for 61\,Vir in
the literature. \citet{1984ApJ...277..640G} determined $B\sqrt{A_{0}}
= 600$\,G\,$\pm 10\%$, where $A_{0}$ is related to $f$ by average
viewing angle. This result is inconsistent with our null detection.
We speculate that the difference between the two measurements is due to
our higher data quality and different, possibly more sophisticated analysis
technique.

\subsubsection{59\,Vir -- HD\,115383}\label{sec:59Vir1Comp}

\begin{figure*}[]
  \centering
  \includegraphics[clip]{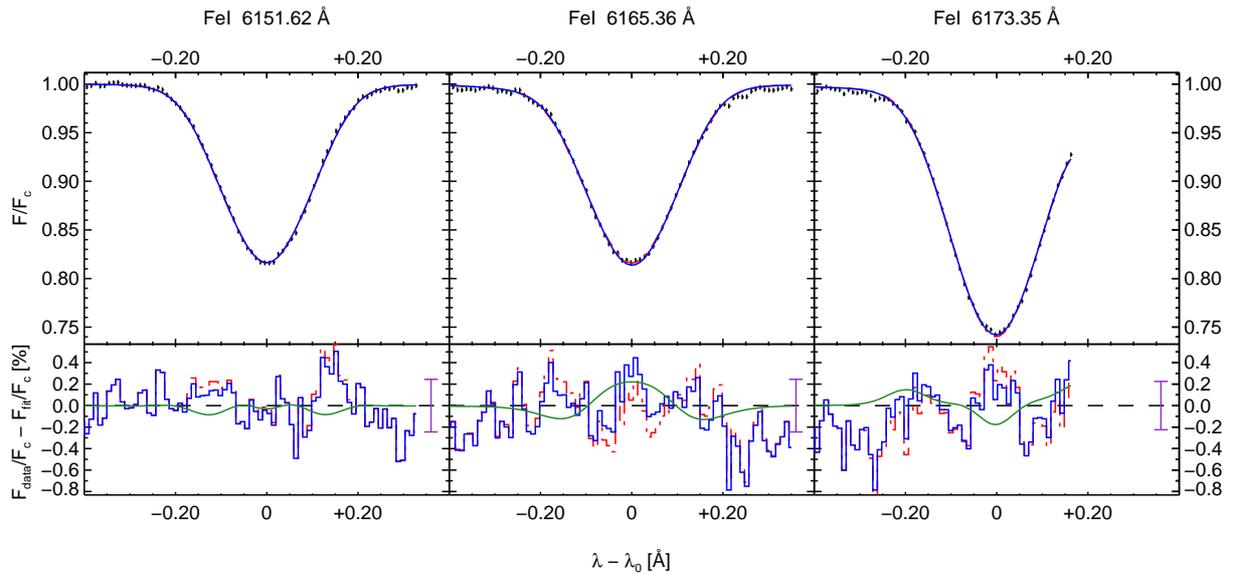}
   \caption{59\,Vir, line set B3, one-component model: Data and
     best-fit with $Bf = 500$\,G (solid blue) and $Bf = 0$\,G
     (dash-dotted red)}
   \label{fig:59VirB3FeI}%
\end{figure*}

59\,Vir is a G0V field dwarf known for its activity and has been
extensively studied by many authors. V\&F05 measured its $v\sin{i} =
7.4$\,km s$^{-1}$, $T_{\rm{eff}} = 6234$\,K, $\log g = 4.60$ and
$\rm{[Fe/H]} = 0.28$ (relative to solar). $\log L_{\rm{X}} = 29.41$
(NEXXUS2), approximately $2.8$\,dex higher than
that of 61\,Vir. Given its spectral type and observed rotation period
$P_{\rm{rot}} = 3.3$\,d measured by \citet{1996ApJ...466..384D},
59\,Vir would be classified as a young star. However, using multiple
age-indicators, \citet{1999A&A...348..897L} favor an age of
$3.8$\,Gyr.  Thus, 59\,Vir is an interesting object to study regarding
the age-activity relationship.

Inversions for 59\,Vir were performed using line sets B3 and B4, cf.
Table\,\ref{tab:atomic}. In the latter case, surface gravity was
well-constrained, despite the less-than-ideal fit to the Ca\,I line,
and $\log g = 4.49$ was obtained, consistent with the
literature value. The other best-fit parameters are also
consistent with the literature
\citep{2005PASJ...57...27T,2008A&A...487..373S}.

We determine $Bf = 500^{+40}_{-60}$\,G for 59\,Vir from line set B3
and $Bf = 525^{+30}_{-60}$\,G from B4, cf. the bottom left and right
panels in Fig.\,\ref{fig:chimaps}. The $3\sigma$ CLs
determined from both data sets are roughly the same with approximately
$300 \le Bf \le 650$\,G. $\chi^{2}$
varies smoothly over the entire $Bf$ range, and a steep incline is
seen above $Bf \sim 800$\,G. Furthermore, the best-fit solution for
$Bf = 0$\,G lies nearly $5\sigma$ from $\chi^{2}_{\rm{min}}$.

The $6173$\,\AA\,line was assigned a slightly higher statistical
weight, since it is the most Zeeman-sensitive line in the set with
$g_{\rm{eff}} = 2.5$. The blue wing of this line contains a blend due
to Eu\,II that was not fully reproduced, despite our effort to
include all available atomic line information in the inversions. Due
to the presence of unidentified blends, the red wing had to be severed
before reaching continuum. Line blends as well as the systematic
signature seen also in the results for the Sun are clearly visible.

The residuals in Fig.\,\ref{fig:59VirB3FeI} indicate improvement in
fit quality due to introduction of magnetic flux for all
lines. However, the largest differences between the magnetic ($Bf =
500$\,G) and non-magnetic best-fit results are of the same
magnitude as the S/N of the data. This result thus illustrates how
difficult it is to detect ZB in optical Stokes $I$, even for
relatively high values of $Bf$. The presence of systematic residual
signatures similar to those found for the Sun and 61\,Vir underline
the need for rigorous and accurate implementation of radiative
transfer and line broadening functions.

\subsubsection{HD\,68456}
The F6 dwarf HD\,68456 was of particular interest in our analysis,
since data were available in both data sets and thus allowed our results to
be tested for consistency. Since the dates of observation were
nearly 4 years apart, we did not simultaneously invert lines from both
sets. After all, changes in atmospheric parameters, for instance due to
activity cycles, within this time span cannot be excluded. Fundamental
parameters of this object are given by \citet{1998A&AS..129..431H} ($T_{\rm{eff}}
= 6396$\,K, $\log \rm{g} = 4.14$, $\rm{[Fe/H]} = -0.29$) and R\&S03
($v\sin{i} = 9.8$\,km s$^{-1}$). NEXXUS2 indicates a high level of
activity of this star ($\log L_{X} = 29.05$). Furthermore, HD\,68456
is a particularly interesting object to study in this context, since
we are not aware of any previous ZB measurement in a late F-type
star\footnote{but note the work of \citet{2008MNRAS.385.1179D} on the
  changing magnetic field topology on $\tau$\,Boo (F7 IV-V
  \citep{2001AJ....121.2148G})}.

Relatively fast rotation and high effective temperature combine to produce
very shallow absorption lines in this star's spectra. This fact complicated
analysis of line set A3 where lines are generally shallower than in
data set B. Furthermore, data set A for this star was of lower quality
than the other spectra analyzed in this work, containing individual
data points that deviate significantly from the general line shape,
while continuum S/N was approximately 400 in both data sets. The
deviations seen are not consistent with the signature expected from a
starspot.

The top and center right panels in Fig.\,\ref{fig:chimaps} show our
results obtained for line sets A3 and B4, respectively. For line set
A3, the overall best-fit $Bf = 1100$\,G, but no detection can be
claimed. This is underlined by an F-test probability $P_{\rm{f}} =
0.21$ for this result indicating spurious fit improvement. Using line
set B4, however, we do find evidence for a magnetic field and obtain
$Bf \sim 1$\,kG, with the formal 
$3\sigma$ CL being  $600 \leq Bf \leq 1200$\,G. 

\subsection{Two-component inversions}
The concept of two-component (TC) inversions was introduced in Sect.
\ref{sec:twocomp}. We carried out these more complex inversions
for the Sun and 59\,Vir. For the Sun, for which no one-component detection of
a magnetic field was made, we did not expect reliable 2-component results.
These inversions were carried out purely as a test. In the illustrative case
discussed in the first subsection below we made the starting assumption
for magnetic flux to be concentrated in cool regions, such as spots. The
results are quite instructive regarding systematic effects.
For 59\,Vir, where the probability of attaining additional information
from a two-component inversion is higher, we investigated the three 
cases mentioned above: equal temperatures for both components; 
warm magnetic component (warm case); cool magnetic component (cool case).

\subsubsection{The Sun}\label{sec:SunTC}
\begin{figure}[b]
  \centering
  \includegraphics[clip]{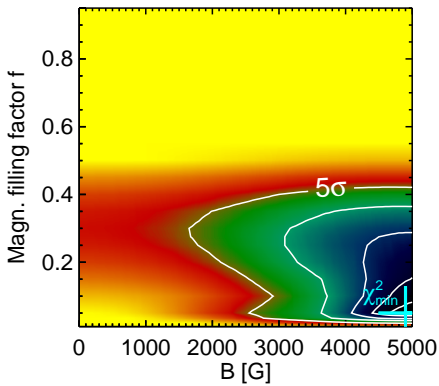}
  \caption{$\chi^{2}$-map for the Sun, line set A3, TC model with cool magnetic regions.}
  \label{fig:Gany_chi2map}
\end{figure}
Figure\,\ref{fig:Gany_chi2map} shows the $\chi^{2}$-map calculated from
TC inversions of solar line set A3. The input atmospheres used were 
\texttt{asun.dat13} (cf. Sect.\,\ref{sec:Gany1comp}) for the warm and 
a $T_{\rm eff} = 4750$\,K Kurucz atmosphere for the cool component. 
We obtain $\chi^{2}_{\rm{min}} = 170$, a value considerably
lower than that obtained for the OC model, for
$B = 4950$\,G and $f = 0.05$. The $Bf$ grid
was calculated for field strengths of up to
$5000$\,G, with filling factors up to $0.95$ for $B \leq 2000$\,G and
$f \leq 0.5$ for $B > 2000$\,G. The CLs are $235 \leq Bf
\leq 250\,{\rm G}\,(1\sigma)$ and $150 \leq Bf \leq 1400\,{\rm G}\,(3\sigma)$,
and thus would suggest evidence for very strong surface fields in
the Sun.
However, there are a number of problems with these results. For
instance, magnetic 
fields in sunspots do not usually exceed $\sim 3$\,kG while the average
field strength of sunspots is roughly $1000 - 1300$\,G
\citep{1993A&A...267..287S} and the field strength averaged over sunspot
umbrae usually does not exceed 2500\,G. 
In contrast, our $3\sigma$ confidence
level suggests that $B$ is well above $4$\,kG with $f$ between $1 - \sim 30\%$
in disagreement with the observed sunspot coverage which is (always)
below $1\%$.
Furthermore, a multitude of measurements places the average
  $Bf$ of the (quiet) Sun considerably lower \citep{2009ASPC..405..135S}.
Last, the $3\sigma$ range for $Bf$ is vastly larger than
that found for the OC result. Therefore, we conclude
that our fit was not well-constrained and that in this case the more 
complex TC model yielded less significant results than the simpler OC model.

\begin{figure*}
\centering
\includegraphics[clip]{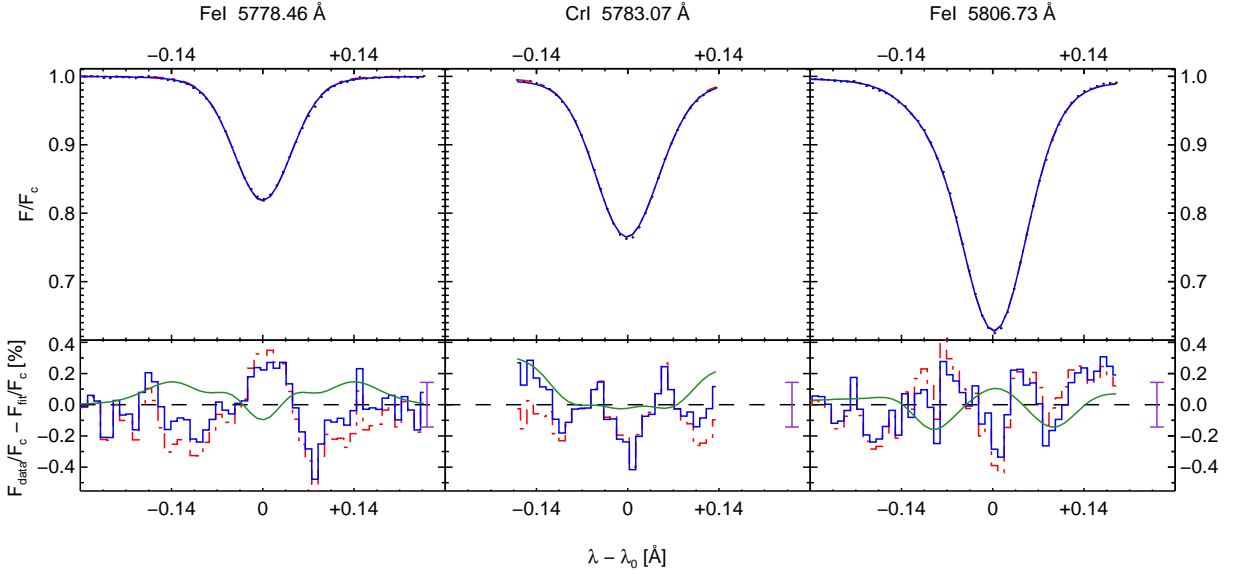}
\caption{Sun, line set A3, comparison of overall best-fit TC ($B =
  4900$\,G, $f = 0.05$, blue solid) and OC ($Bf = 0$\,G,
  red dash-dotted) results.}
\label{fig:Gany_4900.05vs0}
\end{figure*}
Figure\,\ref{fig:Gany_4900.05vs0} helps to understand what happened here.
The greatest decrease in $\chi^{2}$ was
achieved by decreasing the systematic signature in the
$5778$\,\AA\,Fe\,I line. This apparent improvement in fit quality has
stronger influence on $\chi^{2}$ than the deteriorating agreement
between the measured and calculated profiles of the Cr\,I line at
$5783$\,\AA, since the former line contributes more degrees of
freedom. Therefore, the magnetic field (or rather the lower temperature
associated with the magnetic component) improves the fit of the 
Zeeman-insensitive ($g_{\rm{eff}} = 1.2$) line, while the Zeeman-sensitive
($g_{\rm{eff}} = 2.0$) line is less well matched. The worse fit in the
blue Cr line wing is
compensated in part by a small unidentified blend in the red wing that
is falsely removed by the magnetic model. Thus, our conclusion is that
this TC inversion is driven by systematic difficulties and
yields a result that is less representative of the physical processes,
albeit superior in terms of pure $\chi^2$.  It therefore serves as
an example of how line blends and other systematic effects on line
shape affect our results.

\subsubsection{59\,Vir -- HD\,115383}
The results obtained for 59\,Vir  using  OC models indicated formal  
$3\sigma$ detections of significant magnetic flux. To assess, if this
detection is actually robust and significant, we performed TC
inversions for this strongest candidate. The following results were
obtained from our physically motivated TC inversions that
distinguish between magnetic and non-magnetic components.

We investigated three different cases of TC inversions:
equal temperatures, cool magnetic, and warm magnetic components.
Thus, in the latter two cases, we investigate the effect of having
different temperatures for the two components.

For the equal temperature case, we used the $T_{\rm{eff}} = 6250$\,K
Kurucz atmosphere for both components and coupled the two parameters
$T_1$ and $T_2$, forcing them to be equal. In the cool case, we used
tabulated atmospheres calculated for
$T_{\rm{eff}} = 6250$\,K and $5000$\,K for the non-magnetic and the
magnetic component, respectively. The warm case was motivated by
the warmer plage or network regions on the Sun. The Kurucz atmospheres
were those with $T_{\rm{eff}} = 6250$\,K for the magnetic and
$T_{\rm{eff}} = 6000$\,K for the non-magnetic regions. The temperature
parameters $T_1$ and $T_2$ could vary freely and independently in both
the warm and cool cases. The fitting ranges for $T_1$ and $T_2$ were
large enough to allow turning the cool component into the hotter one
and vice versa. The starting values of the fit parameters were
identical in both cases, apart from the inverted values of $T_{1}$ and
$T_{2}$ (the second temperature component
was magnetic). All other starting values were identical
to the OC inversions of $59$\,Vir.

\begin{figure*}
  \subfloat[$\chi^{2}$-map, same atmospheres for magn. and non-magn. regions]{\label{fig:59Vir_chi2map_luke}\includegraphics[clip]{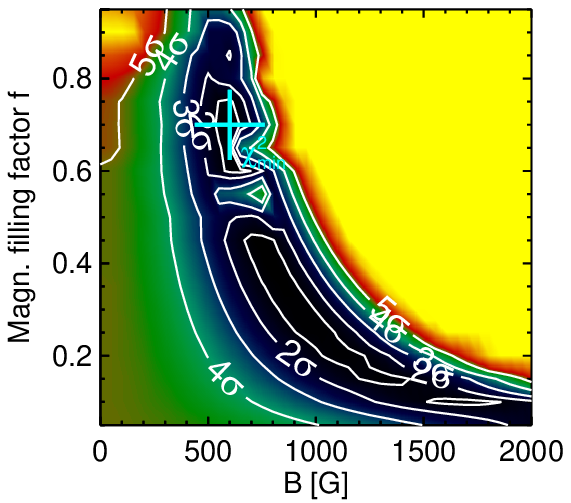}}
  \subfloat[$\chi^{2}$-map, cool magnetic regions]{\label{fig:59Vir_chi2map}\includegraphics[clip]{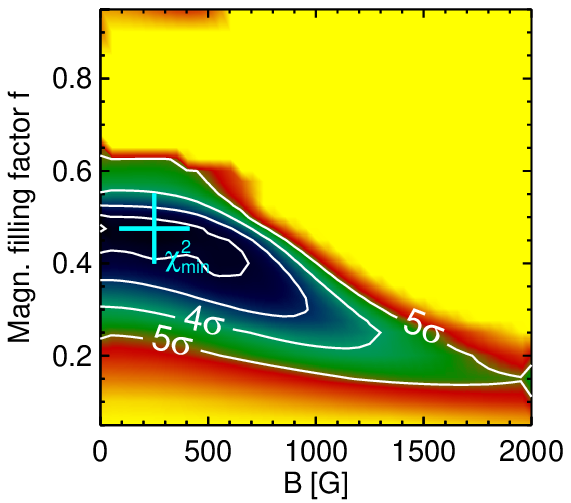}}
  \subfloat[$\chi^{2}$-map, warm magnetic regions]{\label{fig:59Vir_chi2map_hotter}\includegraphics[clip]{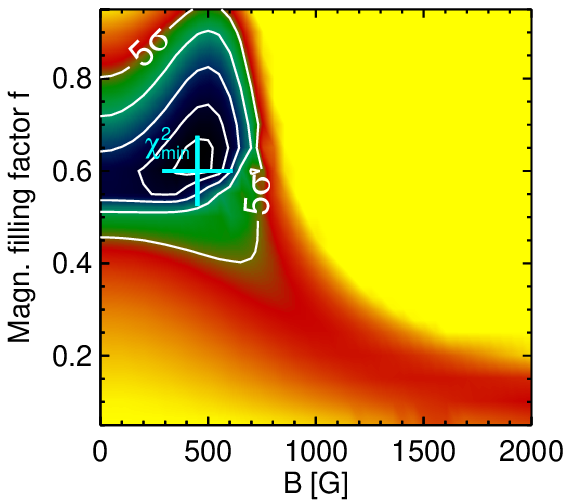}}
  \caption{Results obtained from TC
    inversions of 59\,Vir, line set B3. $\chi^{2}_{\rm{min}}$ indicated by
    large cyan pluses. CLs indicated by solid white lines.}
  \label{fig:chimap2comp}
\end{figure*}

\paragraph{Equal temperatures:} Figure\,\ref{fig:59Vir_chi2map_luke} shows
the $\chi^2$-map obtained from our \label{sec:equal}
inversions of a model with equal temperatures for both components. As can be
seen in Table\,\ref{tab:results}, our overall best-fit parameters are virtually
identical with our OC results. Even the two $Bf$ values lie
  within their 1$\sigma$ CLs. $\chi^2_{\rm{min}}$ is reached for $B =
  700$\,G and $f=0.6$. However, 
even the 1$\sigma$ CL spans a large range in $f$-values, and the
3$\sigma$ CL is consistent with almost any $f$ value.
The reason for this lies in the $\chi^2$ surface, which has the well-known
banana-shape, with a minimum running from the
combination of high $f$ and small $B$ to the opposite extreme of small $f$ and
large $B$, while maintaining roughly equal magnetic flux (i.e. $Bf$).
This is the signature of degeneracy between these two parameters.
Of particular importance is the fact that in spite of the additional parameter
$f$, the  $\chi^2_{\rm{min}}$ is identical to the OC case, which also indicates
that there is not sufficient information in the data to distinguish between
$B$ and $f$. Furthermore, the formal $Bf$ error
margins for this case are larger than for the OC inversions and
the other best-fit parameters nearly identical. Therefore,
we conclude that a distinction made between magnetic and non-magnetic regions
is insignificant.

\citet[hereafter Li94]{1994ASPC...64..438L} determined $Bf = 190$\,G
for 59\,Vir \citep[number published in][]{1996mpsa.conf..367S} using an
equal temperature TC model. Their $Bf$ value is lower than, yet 
consistent with, our $3\sigma$ $Bf$ range obtained from the TC inversions 
in the equal temperature case. It is, however,
inconsistent with our $3\sigma$ $Bf$ range determined from the OC inversions 
that favor higher $Bf$.
At present it is not possible to say whether the difference to our results is
due to differences in technique or due to the evolution of the stellar
magnetic field.

\paragraph{Cool magnetic regions:}\label{sec:cool}
\begin{figure*}
  \centering
  \includegraphics[scale=0.85,clip]{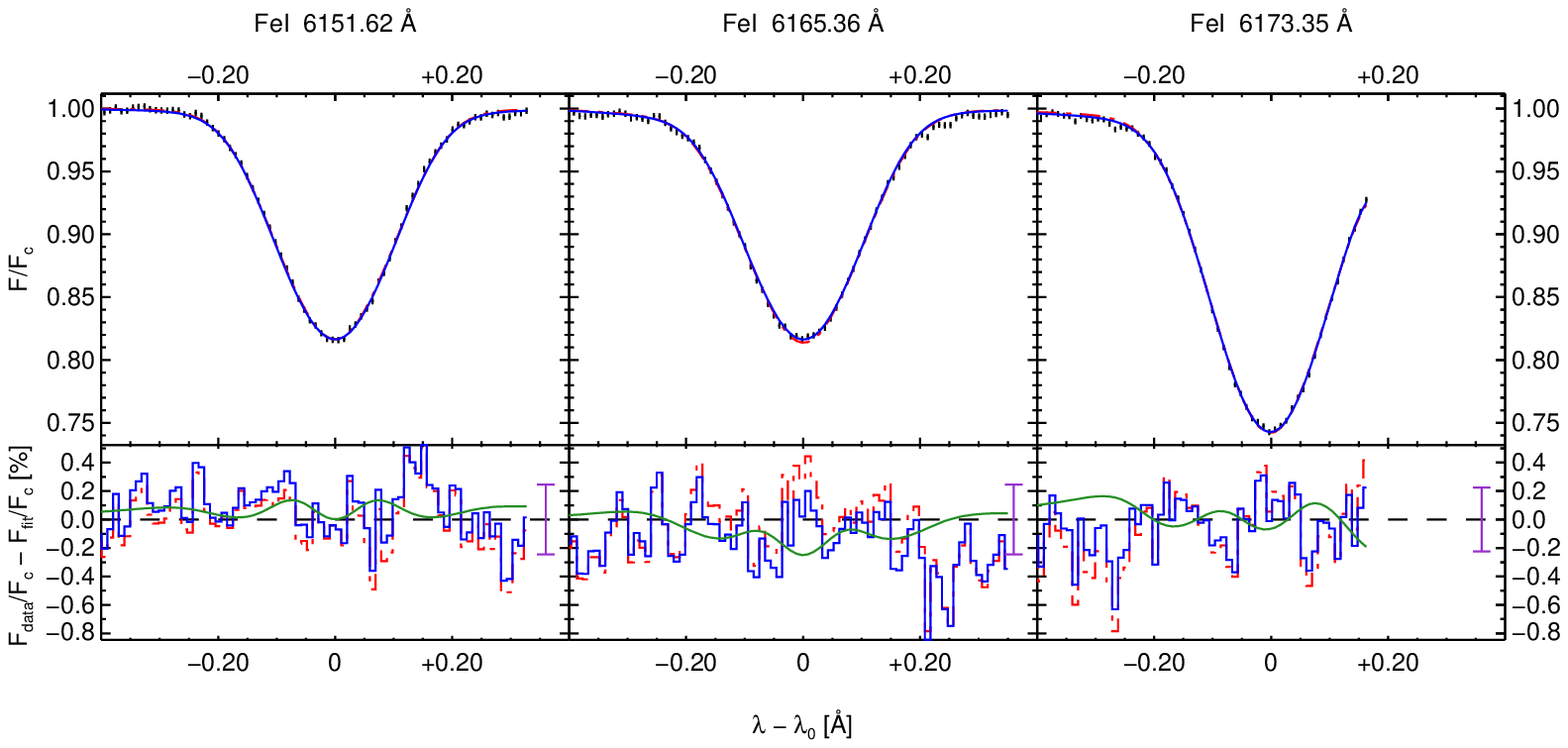}
  \caption{59\,Vir, line set B3: Comparing $Bf = 120$\,G
    TC (blue solid) to $Bf = 500$\,G OC (red
    dash-dotted) best-fit
    results.}
  \label{fig:59Vir_B3_fits_2CompVs1Comp_b}
\end{figure*}
As in the above example for the Sun, we assume the majority of the magnetic
flux to originate in starspots, i.e. in cool regions on the stellar surface.
Figure\,\ref{fig:59Vir_chi2map} shows the $\chi^{2}$-map corresponding
to this case. The overall best-fit $\chi^{2}_{\rm min} = 216$ is
reached for $B = 250 - 300$\,G with $f = 0.475$ for both,
i.e. $Bf = 120 - 140$\,G. $\chi^{2}_{\rm{min}}$ is lowered by
$46.3$ relative to the OC and TC equal temperature models and the formal
$3\sigma$ confidence level is $0 \leq Bf \leq 320$\,G.
The formal $3\sigma$ range in $Bf$ is smaller than in the equal temperature
TC case. The improvement in fit
quality over the OC model is most clearly seen in the better matching
central depth of the Zeeman-insensitive $6151$\,\AA\ and of the Eu\,II
blend to the Zeeman-sensitive line at $6173$\,\AA,
cf. Fig.\,\ref{fig:59Vir_B3_fits_2CompVs1Comp_b}. Both improvements
are likely due to changes in level populations. The presence of a cool
component enables a better matched line depth of Fe\,I, whereas the warm
component contributes additional ionized europium thereby better matching
the Eu\,II blend. In fact, the other two iron lines are also better
matched by the TC inversion.

\begin{figure}[b]
  \centering
\includegraphics[clip]{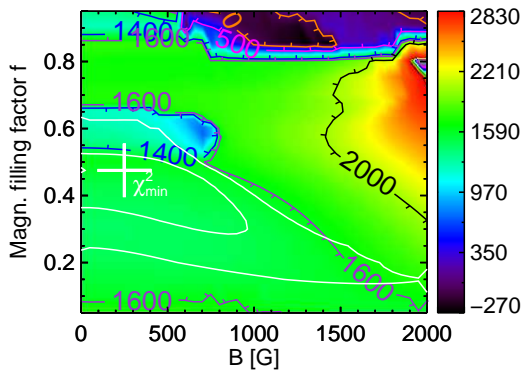}
  \caption{$\Delta T$-map for $59$\,Vir, cool case. 
    Difference in K between $T_1$ and
  $T_2$ indicated by colors indicated in legend on right. 
    Contours drawn for $\Delta T = T_1 - T_2 = 0$\,K (orange), 
    $500$\,K (magenta), $1400$\,K (blue), $1600$\,K (purple), 
    and $2000$\,K (black).
    White contours indicate 3 and 5$\sigma$ CLs. The large white 
  plus indicates the point of overall best fit.}
 \label{fig:59Vir_cool_tmap}
\end{figure}

Figure\,\ref{fig:59Vir_cool_tmap} clearly shows that the difference in 
best-fit $T_{1}$ and $T_{2}$ within the
$3\sigma$ CL lies between $1400 - 1600$\,K. This is
roughly in agreement with the difference in temperatures chosen for the
input atmospheres and the temperature differences between the quiet
solar photosphere and sunspot umbrae.

\paragraph{Warm magnetic regions:} \label{sec:warm}
For this case, we employed a model containing slightly warmer
magnetic than non-magnetic regions, i.e. assumed a model
representative of plage or network-regions. This case is particularly
interesting, since a warm second component contributes more flux than
a cool one. Thus, the effect due to a certain $Bf$ on line shape can be
expected to be stronger.

Figure\,\ref{fig:59Vir_chi2map_hotter} illustrates our
results. $\chi^{2}_{\rm{min}} = 226$ is slightly higher than in the above
cool case and yields $B = 450$\,G and $f = 0.6$. $Bf$ ranges from
$ 210 - 325$\,G in the formal  $1\sigma$ interval, while we find it
consistent with $ 0 - 450$\,G at $3\sigma$.

The $3\sigma$ range in $B$ values
is roughly the same as in the cool case, whereas the $5\sigma$ level
is limited to lower $B$ values. Opposite to the cool case, high-$f$
solutions are favored. Thus, both cases favor a slight majority
of the surface covered by warm
elements. Despite the small ($250$\,K) difference in $T_{\rm{eff}}$ of
the input atmospheres, the difference in best-fit parameters $T_{2}$
and $T_{1}$ within the $3\sigma$ C.L. ranges $1300 - 1400$\,K
($\Delta T = T_2 - T_1$ in the warm case, since $T_2 > T_1$). This 
temperature difference is very similar to that found in the case of 
cool magnetic regions (for the $3\sigma$ CL).

\paragraph{Comparison of TC cases (equal, cool \& warm):}
\label{sec:59HotCool} The equal temperature TC case yields results
in agreement with OC inversions, but wider CLs on $Bf$.
Significantly lower $\chi^2_{\rm{min}}$ is found in the two different
temperature cases, also for the non-magnetic best-fit 
solutions (lowest $\chi^2$ for $Bf = 0$\,G in equal, cool \& warm cases:
$287$, $217$, $231$). Interestingly, the warm and cool TC cases tend towards
virtually identical atmospheric configurations with nearly the
same overall best-fit parameters for
$T_{1}$, $T_{2}$ (though $T_1$ and $T_2$ are switched among the cool
and warm cases), $v\sin i$, $v_{\rm{mac}}$, and $v_{\rm{mic}}$,
cf. Table\,\ref{tab:results}.

This suggests that the true reason for the improved fit to the observations
is the presence of two temperature components. These simulate the
inhomogeneity of the stellar atmosphere produced by starspots, granulation,
oscillations, etc. and possibly also non-LTE effects in the line 
formation, which are not taken into account in the LTE modeling we 
have carried out.

A key difference between the different TC cases is the flux
contribution from the respective magnetic component. Due to this
difference in contrast, different $Bf$ values may be expected. Our
inversions yield comparable $Bf$ for the OC and TC equal
temperature model, while $Bf$
is lower for the cases with different temperatures. Among
these latter two, the warm case (brighter magnetic component) yields a
slightly larger $Bf$. However, both are consistent with no magnetic field
at the $3\sigma$ CL (though the warm case is not consistent with
$Bf = 0$\,G at $2\sigma$).

\section{Discussion}
To assess detectability of ZB for
this kind of analysis, we first discuss the general
uncertainties and sources of error before examining the impact of
model assumptions.

Our analysis is prone to uncertainties from various
sources, such as degeneracies among various line broadening
agents and uncertainties in information used as input (atmospheres, line
data, lines used, etc.).
We attempted to avoid degeneracy among line broadening agents by
allowing all fit parameters to vary freely for given average magnetic
fields. This approach helped to nudge the non-linear
$\chi^2$-minimization routine towards the global minimum. In addition,
it provided a clear visualization (see 
Fig.\ref{fig:chimaps}) of the range of acceptable $Bf$ values, thereby 
providing very solid upper limits on $Bf$ of, say, a few hundred to
$1.5$\,kG depending on the star. 

More spectral  lines could have provided more stringent
constraints on model parameters. Unfortunately, our choice of lines was
severely limited by the small wavelength coverage of the CES spectra. 
Temperature stratification would have been even better constrained by using
lines from multiple ionization stages. In order to achieve the fit quality
presented in this paper, however, we relied on lines from different excitation
potentials. We had found that some combinations of 
lines yielded better fit agreement than others. The reasons for this
empirical fact are likely related to blends; specifically, to blend abundances
(often unconstrained), lack of literature on transition parameters needed for
\texttt{SPINOR}, uncertainties in oscillator strengths or rest-wavelengths,
and unidentified line blends or even relative displacements of blend
wavelengths due to convection. For a literature comparison of the fit 
quality presented here, see \cite{2005A&A...444..549F} who applied 
\texttt{SPINOR} in simultaneous inversions of $14$ lines for $\alpha$ 
Cen A and B, including lines from different ionization stages. Their 
Figs.\,2 \& 3 clearly illustrate the larger 
discrepancy between model and data when a large number of lines is used. 

Small systematic effects (best visible in the spectral
difference between observed and computed lines profiles) complicated
our analysis. The high quality of the data revealed these systematic
shortcomings in reproducing real line profiles. The data for the Sun
was both the highest quality spectrum (in terms of S/N) and one of the
least rotationally broadened. The fact that the TC inversions for the
Sun produced such implausible results may indicate that a limit was
surpassed for 
an adequate representation of line profiles by simple broadening
functions such as the radial-tangential macroturbulence profile. 
However, some of the difficulties in the solar TC inversion were also
related to the lower Zeeman sensitivity of line set A3, in addition to
the low level of solar magnetism and the data-related problems of line
blends, etc. Further candidates for the origin of the systematic
signatures at this level may be related to any of a number of sources,
such as  the implementation of limb-darkening and disk-integration,
temperature stratification, neglect of stellar convection (granulation) and 
oscillations, differential rotation, neglect of non-LTE effects, 
depth-dependent magnetic field strength, scattered light in CES data, 
etc. 
In addition, the temperature gradients of active stars may differ
significantly from those of inactive stars. Thus, the representation
of magnetic regions by ``quiet'' atmospheres may also be too crude an
approximation, since it neglects inhibition of convection and the
influence of magnetic fields on temperature stratification. Thus,
applicability of LTE Kurucz input atmospheres 
could be questioned. Non-LTE effects might also be seen, since the systematic
residual signature appears to be enhanced in hotter stars. The analysis thus
suffers from a multitude of
highly degenerate effects originating in approximations. Most of these
flaws cannot be easily resolved, however, and there are valid reasons
for adopting at least some of these approximations.

Two kinds of fitting strategies were explored:
one-component (OC) and two-component (TC) models,
cf. Sect.\,\ref{sec:FittingStrategies} for details. OC models assume a
homogeneous atmosphere with a single temperature stratification and an
average magnetic field, $Bf$, covering the entire stellar surface.
For the Sun and $61$\,Vir, both stars with low X-ray activity, our
OC results excluded $Bf$ exceeding a few hundred Gauss. OC inversions
further yielded evidence of
$500$\,G to $1$\,kG fields for two faster rotators with
higher X-ray luminosity, namely $59$\,Vir and HD\,68456.
As could be expected from its larger maximum $g_{\rm eff}$ value and
$g_{\rm eff}$ coverage, line set B3 was more sensitive to ZB than A3.

Two-component (TC, cf. Sect.\,\ref{sec:twocomp}) inversions were
performed to distinguish between magnetic and non-magnetic surface
regions. For our solar spectrum in line set A3, we present a case study that
assumed the magnetic
fields to be concentrated in spots, i.e. in significantly cooler
regions. However, the solar TC model was driven by systematic effects
(partly due to weak blends in the line wings) and yielded a result
inconsistent with the OC result and the literature.

For $59$\,Vir, we explored three different cases of TC inversions:
a) equal temperatures for both components; b) a cool magnetic, and c)
a warm magnetic component. We find that the equal temperature TC case is
consistent with the results from the OC inversions. Furthermore, the $Bf$
determined is consistent with the only available literature value.
However, no improvement in $\chi^2$ is seen over the OC model,
while an additional fit
parameter, $f$, was introduced. This is consistent with the well known
result that the product $Bf$ is much better constrained than $B$ and $f$
individually \citep{1990IAUS..138..427S}.

The picture changes, however, when the model takes into account two different
temperatures in the two components, as was done for the cool (spots) and
warm (plage and network regions) TC cases. In fact, a significant
improvement of $\chi^2$ was seen for both of these more complex inversions.
Independent of the input atmospheres used for the two components,
these inversions converged towards a very similar photospheric
model featuring similar coverage ($f \approx
50\%$) by hot and cool features with $\Delta T \approx 1300$\,K. The difference
in $\Delta\chi^2_{\rm{min}}\approx 10$ between the warm and cool case
is probably the result of the a priori choice of input atmospheres.
Magnetic flux did not play a deciding role in the inversions'
ability to converge towards this common overall best-fit model.
Note, however, that the improvement of the fit comes at the expense of
additional free parameters (2 new parameters are introduced compared to the
OC case).

For both the warm and the cool TC case, $Bf$ was lower than in the
OC or TC equal temperature case and consistent with no magnetic field
at the $3\sigma$ CL (however, not at $2\sigma$ in the warm case). Thus, while
some evidence for ZB was found in these cases, no detection
can be claimed when allowing for two different temperature components,
one magnetic, the other field free.

Given the significant improvement of $\chi^2_{\rm{min}}$, we find that
model assumptions impact our ability to reproduce the observed line
shape more strongly than the presence or absence of ZB
in optical Stokes $I$, even for relatively high Land\'e factors.
At the same time, the presence of ZB is not excluded, either.

As a next step, even more complex models might be invoked. However,
this does not appear to be expedient at this point, given the
limited number of lines used for analysis, the approximative treatment
of line profiles (e.g. the radial-tangential macroturbulence and the
height-independent microturbulence), the smallness of the Zeeman
signature, and the fact that the TC inversions of $59$\,Vir already
reached the noise limit.

A more prudent approach would be to increase sensitivity to ZB. This
could be done by using high-resolution infrared spectra. In this way,
the $\lambda^2$ dependence of Eq.\,\ref{eq:ZeemanBroadening} could be
exploited to provide stringent constraints that are in addition to the
differential $Bf$-response that was at the base for this investigation. 
Furthermore, the IR contrast between cool and warm regions
would be lower, thereby enhancing Zeeman signatures 
\citep[e.g.][]{2007ApJ...664..975J}.  In
M-dwarfs, successful measurements of strong magnetic fields have
been reported using molecular lines, e.g. using FeH lines in the
Wing-Ford bands \citep{2006ApJ...644..497R,2007ApJ...656.1121R}.
With the majority
of their flux in the infrared and strong fields, these stars
constitute prime candidates for direct $Bf$ measurements using
spectral line inversion as outlined in this work.


\section{Summary}
In this paper, we investigated detectability of ZB in
optical Stokes $I$ spectra of the Sun and three slowly rotating
sun-like stars by performing spectral line inversion with
\texttt{SPINOR}.

Thanks to the ability of
\texttt{SPINOR} to perform spectral line inversion, different
scenarios could be compared and model assumptions evaluated by
scanning the $\chi^2$ plane for fixed $Bf$, or combinations of
$B$ and $f$, values.
This approach enabled us to investigate the form of $\chi^2$ as a
function of $Bf$ as shown in Fig.\,\ref{fig:chimaps} and therefrom
determine formal errors on $Bf$. 

Systematic uncertainties in our analysis are in part related to
literature used as input, such as atomic line data and elemental
abundances. Due to the small magnitude of the Zeeman broadening even
weak line blends complicated the analysis. Approximations, such as the
representation of turbulence in the stellar atmosphere by radial-tangential
macroturbulence and height-independent microturbulence, also influenced
our results.

OC inversions excluded the presence of significant magnetic flux
($Bf < 150$\,G at the $3\sigma$ C.L.) for the
Sun and 61\,Vir. Our results for HD\,68456 may be interpreted as first
evidence for a magnetic field in a late F-type dwarf obtained by
the ZB technique, with $Bf \approx 1$\,kG and formal $3\sigma$
errors of approximately $350$\,G. The OC result for $59$\,Vir is $Bf =
500^{+40}_{-60}$\,G. 

The TC equal temperature case yielded $Bf = 420$\,G for $59$\,Vir, a
higher value than that reported by Li94, but consistent within the
(large) formal $1\sigma$ error. TC inversions with components of
different temperatures were consistent with $Bf$ between $0$\,G and
$450$\,G at the $3\sigma$ CL. The case of a warm
magnetic component was inconsistent with an absence of magnetic field at
$2\sigma$. Previous investigations of this kind frequently used models
similar to our equal temperature TC case. Our investigation of the
different TC inversion cases demonstrates the strong influence of
model assumptions. In fact, the freedom to freely and independently 
vary both atmospheric temperatures in TC models was shown to influence
line shape more strongly than ZB for the lines used in this work,
cf. Sect.\,\ref{sec:59HotCool}. The improvement of $\chi^2$ due to a second
component of different temperature was significantly larger than
that due to a magnetic field, irrespective of whether the magnetic
component was defined as warm, or cool. The present data quality
exceeds that of previous studies, while our sensitivity to Zeeman
broadening as expressed by the $g_{\rm{eff}}$ values in our line sets
is comparable to that in the literature with the exception of studies
in the infrared. 

We conclude that measurements of ZB in optical
Stokes\,$I$ data of slowly rotating sun-like stars are subject
to large uncertainties. These are mostly due to data related
uncertainties, assumptions on atmospheric models, approximations in
line broadening agents, and degeneracy between line
broadening agents. However, the Zeeman sensitivity of every 
individual line (as expressed by its $g_{\rm eff}$ value) 
differs from line to line by a factor of up to $2.5$. This provides a
powerful discriminant for our analysis, since the other relevant
broadening effects depend mostly on temperature or wavelength and are
thus essentially the same for all lines used. It is this differential
reaction of each spectral line to a given magnetic field that ensured
the significance of the measurements presented in this work, despite
the small magnitude of the ZB signature.

The most promising way to improve the present approach is to further
exploit the $\lambda^2$-dependence of ZB and use longer
wavelength data. In the (near) infrared, Zeeman broadening clearly
dominates over Doppler effects and Zeeman splitting can be seen
directly, see e.g.
\citet{1994IAUS..154..437S,1996ApJ...459L..95J,2001ASPC..248..179V}.
New high-resolution near infrared spectrographs have become available for 
these tasks. 


\begin{acknowledgements}
  We thank the referee John D. Landstreet for his detailed report that
  resulted in a much clearer explanation of our analysis and
  a significantly better manuscript.\\
  Thanks are due to the following people: Christian Schr\"oder for
  observing data set B; A. Lagg at MPS for \texttt{SPINOR} related
  support;
  S.H.~Saar for very fruitful advice and discussion.\\
  RIA and AR acknowledge research funding from the DFG under an Emmy
  Noether Fellowship (RE 1664/4-1). RIA further acknowledges
  funding by the Fonds National Suisse de la Recherche Scientifique
  (FNRS).
  The work of SKS has been partially supported by the
  WCU grant No. R31-10016 funded by the Korean Ministry of Education, Science
  and Technology.\\
  This research has made use of NASA's Astrophysics Data System
  Bibliographic Services and the SIMBAD database, operated at CDS,
  Strasbourg, France.
\end{acknowledgements}

\bibliographystyle{aa}
\bibliography{14769}


\Online

\begin{appendix}
\section{Online figures}
We present additional figures illustrating our OC results in
this appendix.

\subsection{Additional one-component fit results}
\begin{figure*}
  \centering
  \subfloat[61\,Vir, line set B3. Blue solid: $0$\,G. Red
  dash-dotted: $175$\,G]{\label{fig:61Vir_B3_1Comp}\includegraphics[scale=0.85,clip]{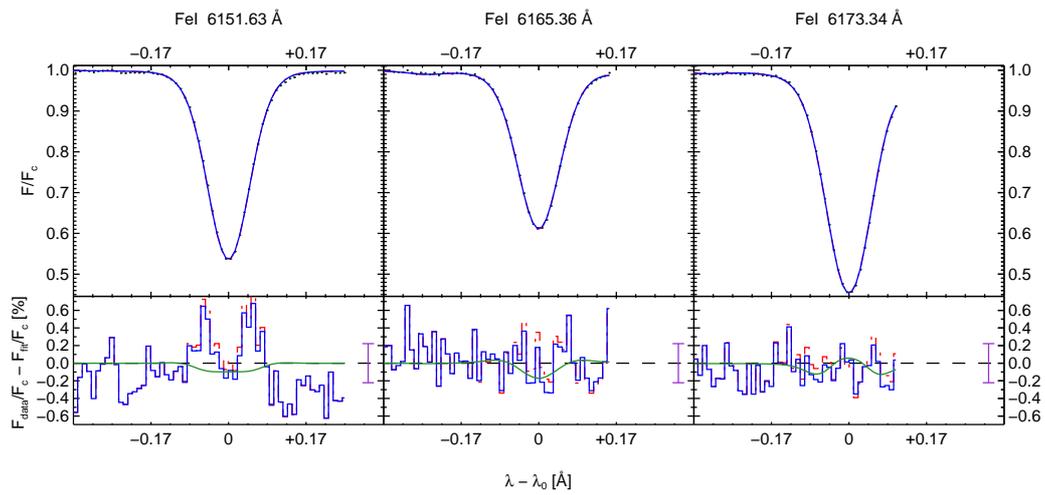}}\\
  \subfloat[59\,Vir, line set B4. Blue solid: $500$\,G. Red
  dash-dotted: $0$\,G]{\label{fig:59Vir_B4_1Comp}\includegraphics[scale=0.85,clip]{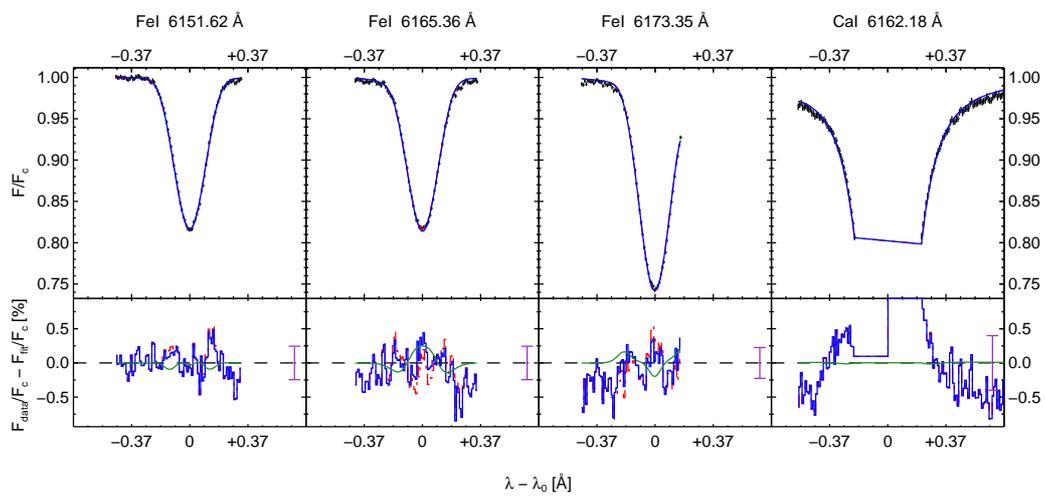}}\\
  \subfloat[HD\,68456, line set A3. Blue solid: $1100$\,G. Red dash-dotted: $0$\,G]{\label{fig:HD68456_A3_1Comp}\includegraphics[scale=0.85,clip]{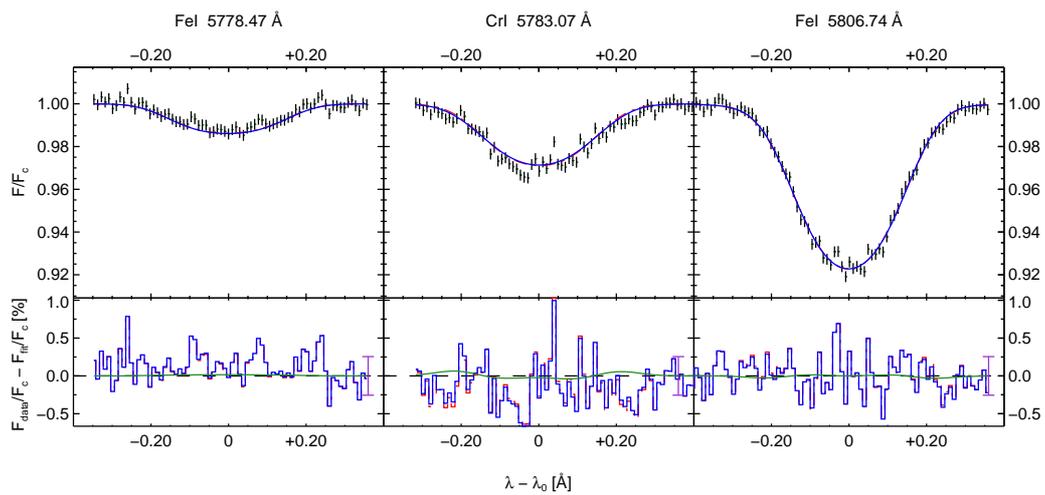}}
  \caption{Additional comparison of overall best-fit results from
  one-component inversions with non-magnetic models and
  $3\sigma$ upper limits, respectively.}
  \label{fig:additionalOCfits}
\end{figure*}

Figure\,\ref{fig:additionalOCfits} shows the OC fit
results not presented in the main article body whose corresponding
$\chi^2$-plots were included in Fig.\,\ref{fig:chimaps}.
The best-fit models drawn are mentioned in the respective captions of
the sub-figures.

\subsection{Free fit parameters}
\begin{figure*}
  \centering
  \subfloat[Sun, line set A3]{\label{fig:Gany_A3_1Comp_parms}\includegraphics[clip]{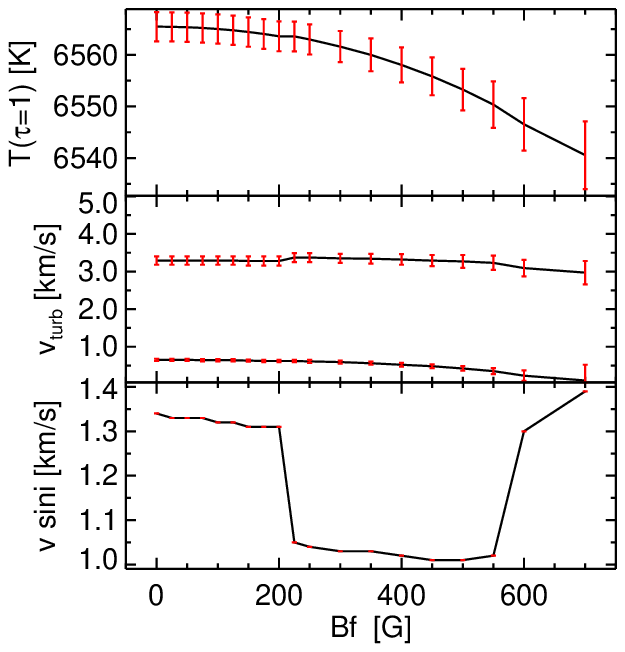}}
  \subfloat[HD\,68456, line set A3]{\label{fig:HD68456_A3_1Comp_parms}\includegraphics[clip]{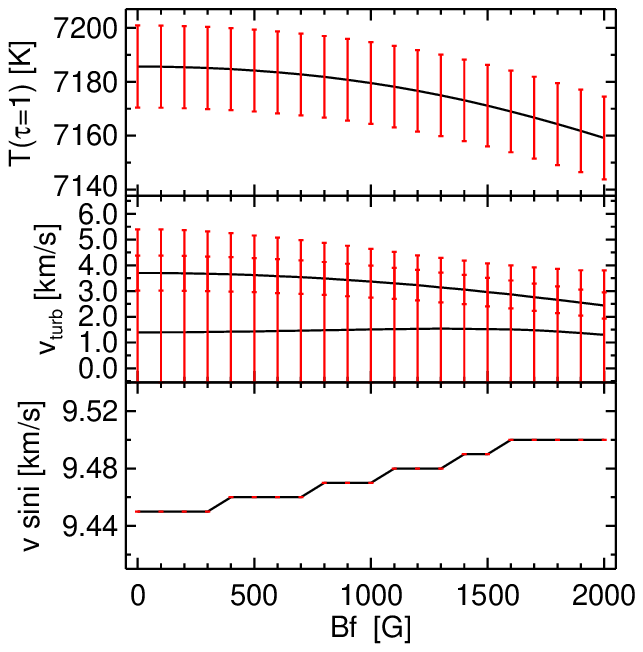}}\\
  \subfloat[61\,Vir, line set B3]{\label{fig:61Vir_B3_1Comp_parms}\includegraphics[clip]{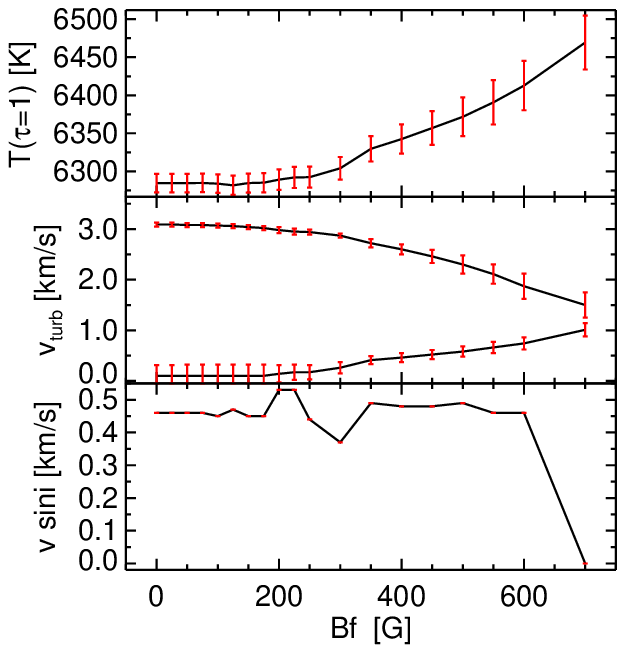}}
  \subfloat[HD\,68456, line set B4]{\label{fig:HD68456_B4_1Comp_parms}\includegraphics[clip]{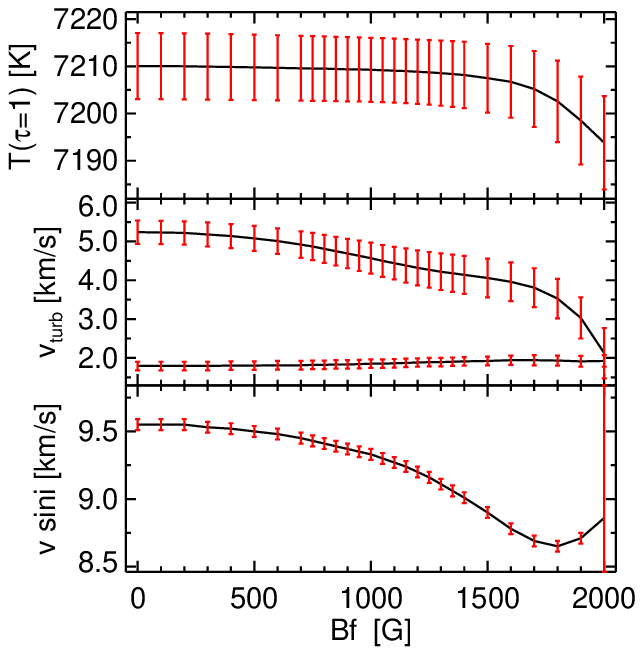}}\\
  \subfloat[59\,Vir, line set B3]{\label{fig:59Vir_B3_1Comp_parms}\includegraphics[clip]{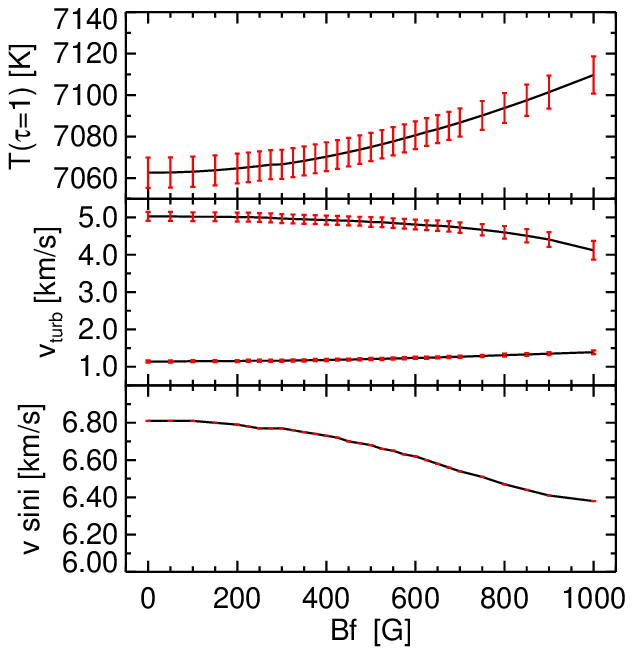}}
  \subfloat[59\,Vir, line set B4]{\label{fig:59Vir_B4_1Comp_parms}\includegraphics[clip]{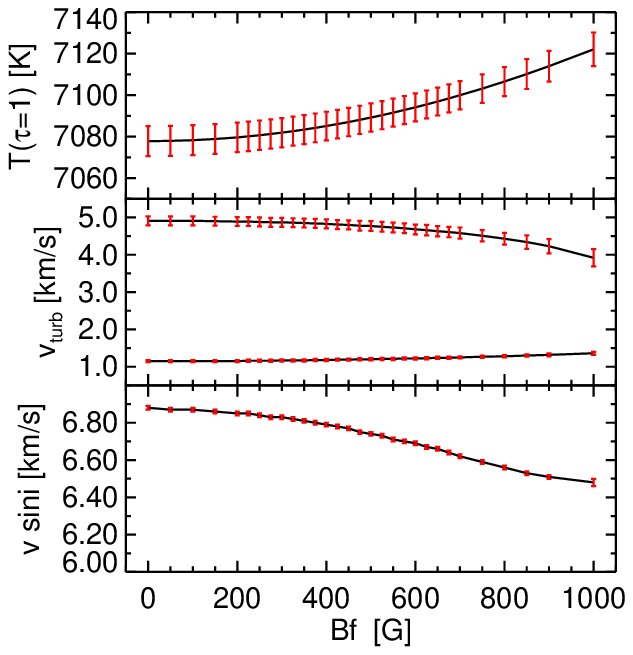}}
  \caption{Behavior of best-fit parameters with $Bf$ using one-component
  models. Plot organized the same way as Fig.\,\ref{fig:chimaps}. v$_{turb}$ indicates micro- and macroturbulent velocities, with macroturbulence usually larger than microturbulence.}
  \label{fig:fitparms}
\end{figure*}

Figure\,\ref{fig:fitparms} shows the behavior of the free fit parameters
with $Bf$ in the OC inversions. Red error bars indicate formal
singular value decomposition (SVD) errors obtained from
\texttt{SPINOR}. They also indicate the
spacing of the $Bf$ grid scanned, thereby illustrating the higher
sampling close to $\chi^{2}_{\rm{min}}$. Generally, fit parameters
behave smoothly, apart from regions where small numerical jumps are
visible, such as for $v\sin i$ in
Fig.\,\ref{fig:Gany_A3_1Comp_parms}. The center panel in each sub-figure 
labeled v$_{turb}$ contains both macro- and microturbulent velocities, 
with macroturbulence generally stronger than microturbulence. Degeneracy 
of fit parameters is
evident from the combined reaction of the parameter set caused by the
fixed values of $Bf$. 

\end{appendix}

\end{document}